\RequirePackage{fix-cm}
\documentclass[twocolumn]{svjour3}          % twocolumn
\smartqed  % flush right qed marks, e.g. at end of proof

\usepackage[utf8]{inputenc}
\usepackage[english]{babel}
\usepackage{appendix}
\usepackage{indentfirst}

\usepackage[backend=biber,
style=numeric,
]{biblatex}
\usepackage{graphicx}
\usepackage{hyperref}
\hypersetup{
    colorlinks=true,
    linkcolor=blue,
    filecolor=magenta,      
    urlcolor=cyan,
}
\hypersetup{colorlinks,linkcolor={blue},citecolor={blue},urlcolor={blue}}

\newcommand{\CA}[1]{\textcolor{black}{#1}}
\usepackage[normalem]{ulem}

\begin{document}
\title{Beirut explosion }

\subtitle{TNT equivalence from the fireball evolution in the first 170 milliseconds}


\author{Charles J.Aouad$^1$        \and
        Wissam Chemissany$^2$  \and
        Paolo Mazzali$^{1,3}$  \and
        Yehia Temsah $^{4}$ \and
        Ali Jahami $^{4}$
        %etc.
}


\institute{Charles Aouad \at 
                \\
              \email{charlesaouad@aascid.ae}           %  \\
           \and
              $^1$ Astrophysics Research Institute, Liverpool John Moores University, IC2, Liverpool Science Park, 146  Brownlow Hill, Liverpool L3 5RF, UK\\
               \and
              $^2$Institute for Quantum Information and Matter, California Institute of Technology,
1200 E California Blvd, Pasadena, CA 91125, USA\\
             \and
            $^3$ Max-Planck Institut fur Astrophysik, Karl-Schwarzschild-Str. 1, D-85748 Garching, Germany \\
            \and
            $^4$ Faculty of engineering, Beirut Arab University, Beirut, Lebanon.\\
}
\date{Received: date / Accepted: date}
\maketitle
\begin{abstract}
The evolution of the fireball resulting from the August 2020 Beirut explosion is traced using amateur footage taken during the first 400 ms after the detonation. 39 frames separated by 16.66 - 33.33 ms are extracted from six different videos located precisely on the map. Measurements of the time evolution of the radius of the shock wave are traced by the fireball at consecutive time sequence \CA{until about $ t \approx$ 170 ms and a distance $ d \approx$ 128 meters}. Pixel scales for the videos are calibrated by de-projecting the existing grains silos building  for which accurate \CA{as built} drawings are available \CA{using the length, the width and the height} and by defining the line of sight incident angles. \CA{In the distance range $ d \approx$  60-128 m from the explosion center, the evolution of the fireball follows a Sedov-Taylor model with spherical geometry and an almost instantaneous energy release. This model is used to derive the energy available to drive the shock front at early times. Additionally, a drag model is fitted to the fireball evolution until its stopping distance at time $ t \approx 500$ ms at distance $d \approx$ 145$\pm$ 5 m. Using the derived TNT equivalent yield, the scaled stopping distance reached by the fireball and the shock wave-fireball detachment epoch within which the fire ball is used to measure the shock wave are in excellent agreement with other experimental data. A total TNT equivalence of $ 200\pm 80$ tons at a distance  $ d \approx 130$ m is found. Finally, the dimensions of the crater size taken from a hydro-graphic survey conducted 6 days after the explosion are scaled with known correlation equations yielding a close range of results. A recent published article by \textcite{dewey2021tnt} shows that the Beirut explosion TNT equivalence is an increasing function of distance. The results of the current paper are quantitatively in excellent agreement with this finding.} These results present an argument that the actual mass of ammonium nitrate that participated in the detonation is much less than the quantity that was officially claimed available.
\end{abstract}

%\begin{multicols}{2}

\section{Introduction}
\label{sec:1}
On the 4th of August 2020, an explosion occurred in the port of Beirut, Lebanon, after a fire ignited in warehouse number 12. This tragic event resulted in massive large scale destruction, severe damage to buildings in an extended radius around the center and loss of lives. It was claimed by officials that this hangar contained an amount of 2750 tons of ammonium nitrate kept in the port for around 6 years.
\par A few attempts have already been made to quantify the amplitude of this explosion.
\CA{Several of these studies use the time of arrival of the shock wave (from audio and visual inspection of footage) up to distances ranging from 500 to about 2000 m  \parencite{rigby2020preliminary,stennett2020estimate,pasman2020beirut}. These studies use empirical relations linking the scaled time of arrival with the scaled distance to report a TNT equivalence range of 0.3-1.1 kt. \textcite{pilger2020yieldseismic} use open access seismic data to yield a range of 0.5-1 kt TNT.}
  \par \textcite{diaz2020explosion} measures the evolution of the fireball until about 200 m distance from the center. He uses 26 data points taken from publicly available videos to yield a range of 0.5-0.6 kt of TNT.
 The data of \textcite{rigby2020preliminary} is used by \textcite{dewey2021tnt} to compare the peak hydrostatic overpressures with experimental measurements for both TNT and ANFO explosions. He finds that Beirut explosion produces overpressures that are weaker than overpessures that would have been produced by a TNT explosion of the same energy at close distances to the center, while at larger distances it produces slightly larger over-pressures. He concludes that the TNT  equivalence of the Beirut explosion is an increasing function of distance (refer to Fig 3 in \cite{dewey2021tnt}).
 \par In this paper we report the TNT equivalence by measuring the kinematics of the fireball in the close proximity of the center (in a distance range of 60-145 m from the explosion center in the first 170 ms). We base our study on both experimental observations of the fireball evolution generated by chemical explosions as reported by \textcite{gordon2013fireball}, and using a Sedov-Taylor model to derive our results \parencite{taylor1950formationI,taylor1950formationII,sedov1946propagation}.
\par The paper is divided as follows: in section \ref{sec:2}, we present our methodology, in section \ref{sec:3}, we present our results and observations and we discuss these results. In section \ref{sec:4}, we compare our work to the literature and in section \ref{sec:5} we present some additional thoughts. We draw our conclusions and discuss possible lines of future work in section \ref{sec:6}.
%%%%%%%%%
%%%%%%%%%%
\section{Methodology}
\label{sec:2}

\subsection{Background}
\label{subsec:1}
\setlength{\parindent}{6ex} Explosions are the swift release of a large amount of energy \parencite{kingery1984airblast,lewis2012combustion,baum1959physics}. This process is usually caused by the ignition of a `fuel'. A burning front is then formed and propagates within the medium burning it as it proceeds. The power of this process: i.e. the energy release per unit time, depends not only on the chemical or nuclear potential of the fuel, but also on the velocity of propagation of this burning front throughout the material. In a pure deflagration, this burning front propagates subsonically and causes burning by heat transfer \parencite{kadowaki1995deflagration,nomoto1976carbondeflagration}. 
\par A more powerful form of burning is a detonation, in which case, the burning front travels supersonically creating a shock front ahead of it and can cause the burning of the fuel by compressive heating \parencite{woosley2007type,mazzali2007detonation}. These energetic burning explosive phenomena are observed in nuclear and in chemical reactions on a wide range of magnitudes ranging from small controlled industrial activities to large astrophysical contexts (e.g. solar flares, supernovae). It is usually accepted that a deflagration can transit to a detonation in suitable conditions \parencite{liberman2010deflagration,khokhlov1999numerical,oran2007origins}, although this transition is still an extensive theoretical area of research. Once the shock wave reaches the outer boundary of the burning fuel, it will be transmitted to the surrounding medium (whether a fluid or a solid) and will propagate isotropically in the form of a blast wave.
\par This sudden increase in pressure will cause the ignited hot material and the gaseous residues of the reaction to expand rapidly with high velocities pushing on the fluid around them \parencite{kato2006numerical,kwak2011expanding}. Furthermore, the sudden increase in pressures at the vicinity of the detonation will cause an increase in temperature (few thousands degree kelvin). Finally the transfer of momentum between the shock front and solid particles will also cause these solid particles to accelerate spherically away from the center \parencite{zhang2003shock, jenkins2013explosively, rigby2020reflected}. The combined effect will result in the creation of an optically \CA{thick} visible fireball.
\CA{Immediately after the explosion, both the fireball and the shock front rapidly expand. However, the expansion of the fireball will decelerate until it halts and reaches a stopping distance while the shock front will detach and keep expanding depositing energy until it decays into a sonic wave. As the rate of fireball expansion decreases, both temperatures and atmospheric over-pressures will also drop making its sharp edge less defined and less luminous \parencite[see also][sect303-304]{united1973nato}  \parencite{gordon2013fireball,bethe1958blast,taylor1950formationII,sedov1946propagation}.}
\
%%%%%%

\subsection{Mathematical description}
\label{subsec:2}
\CA{The theoretical models for the study of the shock wave generation and propagation were developed by studying nuclear explosions \parencite{bethe1958blast,sedov1946propagation,taylor1950formationI}.} \textcite{taylor1950formationI} considered the solution where the energy is released instantaneously, in a very small volume (point source) and where the mass of the explosives is insignificant: i.e. nuclear explosion. He then derived the evolution of the shock wave by solving numerically three differential equations, namely an equation of motion, a continuity equation and an equation of gas state taking the boundary conditions given by the Rankine-Hugoniot equations \parencite{rankine1870xv,hugoniot1887memoir}. Later, he was able to experiment the accuracy of his work after the publication of the photographs of the Trinity nuclear explosion test \parencite{szasz1984daytrinitytest,mack1946trinityphotos}. 
\CA{Although Taylor analysis describes best a nuclear explosion, he later provided evidence, based on experimental data, and showed the boundaries (scaled distance) where a chemical explosion can resemble a nuclear explosion and where his model can still be applied \parencite{taylor1950formationII}.}

The Taylor model has the form: 
\begin{equation}
R^5=t^2 E K^{-1} \rho^{-1}_0.
 \label{eq1}
\end{equation}

Here $\rho_0$ is the undisturbed gas density, $K$ is a constant that depends on $\gamma$ the ratio of specific heats of the gas and $E$ the part of the energy that has not been radiated away.\\
This relation can be written as
\begin{equation} 
\frac{5}{2}\log_{10}R=a \log_{10}t +b
 \label{eq2}
\end{equation}
where $a$ is the slope of this linear relation and is expected to be equal to unity if the observation follows the theoretical prediction. In that case, the energy $E$ can be calculated from Equations \ref{eq1} and \ref{eq2} as: 
\begin{equation} 
 E = 10^{2b} K \rho_o
 \label{eq3}
\end{equation}
where 
\begin{equation} 
 10^{2b} = R^5 t^{-2}.
 \label{eq4}
\end{equation}
the term $b$ can be calculated from a linear fitting to the observed data \parencite{taylor1950formationII}.

\CA{\textcite{sedov1946propagation} developed a more general form which depends on the geometry of the explosion and on the rate of energy release, this model is also a power law whose exponent depends on the dimensionality of the event and on the rate of energy release. It has the form of: }
\CA{\begin{equation}
R_\mathrm{s}(t)=a t^b,
 \label{eq5}
\end{equation}}
 where 
\CA{\begin{equation}
\label{eq6}
b=\frac{s+2}{n+2}.
\end{equation}}

\CA{Here $n$ is the dimensionality of the expansion: $n=1$ for a planar expansion, $n=2$ for cylindrical and $n=3$ for spherical, $s$ is a factor describing the rate of energy release: $s=0$ for instantaneous energy release and $s=1$ for continuous energy release. 
In the case where $n=3$ and $s=0$ the Sedov solution will be similar to Taylor solution: i.e. a power law with an exponent of 0.4.}  

 \CA{The applicability of the Sedov-Taylor model is valid in the region where the shock wave has expanded to displace a mass of air exceeding the explosives mass and where the pressure differential across the shock is high compared to the ambient background \parencite{gordon2013fireball}. Immediately after the explosion, the location of the shock wave radius can be traced by the fireball before these two separate. This has been demonstrated before \parencite{taylor1950formationII, gordon2013fireball,rigby2020reflected,rigby2020preliminary,jenkins2013explosively,dewey2021tnt,diaz2020explosion,zhang2003shock}. Therefore, tracing the fireball kinematics can trace the shock wave and the blast energy can be calculated accordingly.}

\begin{figure}[ht]\centering
  \includegraphics[width=0.45\textwidth]{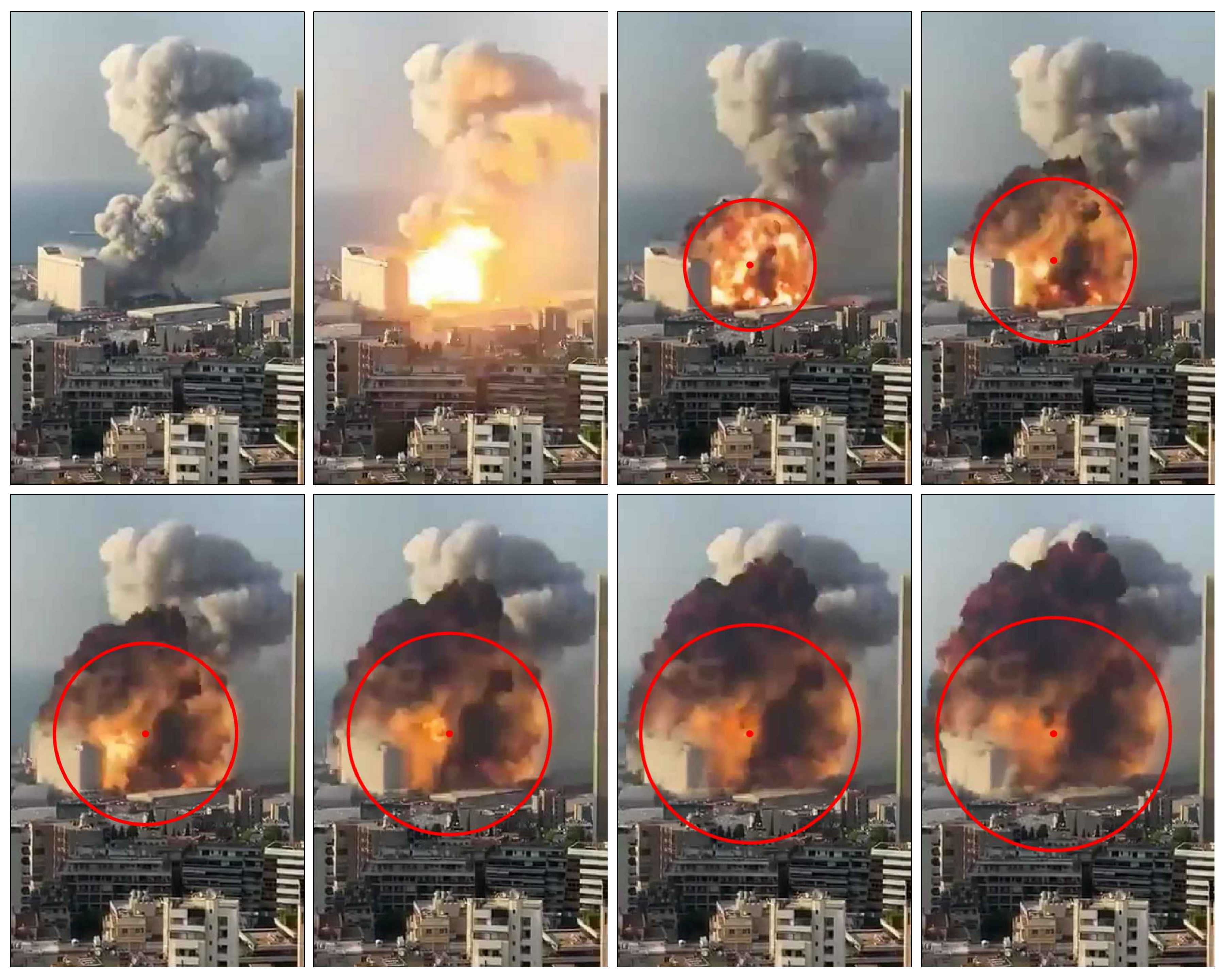}
\caption{Video 2 taken from 1400 m distance. 8 frames separated by 33.33 ms showing the fireball along with the circle fit. The detonation is assumed to have happened anytime between the first and second frame.}
\label{fig:1}       
\end{figure}
\begin{table}[ht]

\caption{Videos used to extract frames. Column $1$ shows the video label, column $2$ shows the frame rate in frames per second (FPS) for each video. Column $3$ shows the distance in meters, column $4$ shows the incident line of sight in degrees. The last column shows the video reference as taken from social media.}
%\caption{Videos used to extract frames }
    
%
\small\addtolength{\tabcolsep}{-4pt}
\scalebox{0.91}{
\begin{tabular}{lllll}
\hline\noalign{\smallskip}
video &  rate & distance & $\alpha$ Ref \\
 label & FPS& meters   &  $^{\circ}$ &  \\
  
\noalign{\smallskip}\hline\noalign{\smallskip}
1 & 30 & 1146 & 83 \parencite[][Tweet]{Ghattas}\\
2 & 30 & 1400 & 22 \parencite[][Youtube]{arabiya}\\
3 & 30 & 550 & 24 \parencite[][Youtube]{arabiya}\\
4 & 30 & 666 & 14 \parencite[][Tweet]{CN}\\
5 & 60 & 630 & NA \parencite[][Youtube]{lanacion}\\
6 & 30 & 1026 & 24 \parencite[][Youtube]{nm}\\

\noalign{\smallskip}\hline
\end{tabular}}

\label{tab:1}  

\end{table}
%%%%%%
%%%%%%%
\begin{table}[ht]\centering

%\caption{Fireball evolution in the first 230 ms}
%\label{tab:2}       % Give a unique label
\begin{tabular}{lllll}
\hline\noalign{\smallskip}
R & pixel scale $\theta$ & R & T & video \\
pixels  & $\mathrm{m}/\mathrm{pixel}$   &   meters &   seconds\\
  
\noalign{\smallskip}\hline\noalign{\smallskip}
58 & 1.373 & 79.634 & 0.05 & 1 \\
72 & 1.373 & 98.856  & 0.083 & 1 \\
82 & 1.373 & 112.586 & 0.116 & 1 \\
90 & 1.373 & 123.570 & 0.149 & 1 \\
97 & 1.373 & 133.181 & 0.183 & 1 \\
102 & 1.373 & 140.046 & 0.216 & 1 \\
71 & 1.12 & 79.52 & 0.05 & 2 \\
89 & 1.12 & 99.68 & 0.083 & 2 \\
99 & 1.12 & 110.88 & 0.116 & 2 \\
110 & 1.12 & 123.20 & 0.149 & 2 \\
119 & 1.12 & 133.28 & 0.183 & 2 \\
127 & 1.12 & 142.24 & 0.216 & 2 \\
189 & 0.3443 & 64.827 & 0.033 & 3 \\
255 & 0.343 & 87.465 & 0.066 & 3 \\
300 & 0.343 & 102.900 & 0.099 & 3 \\
332 & 0.343 & 113.876 & 0.133 & 3 \\
62 & 1.105 & 68.510 & 0.033 & 4 \\
70 & 1.105 & 77.350 & 0.050 & 4 \\
83 & 1.105 & 91.715& 0.066 & 4 \\
91 & 1.105 & 100.555 & 0.083 & 4 \\
97 & 1.105 & 107.185 & 0.100 & 4 \\
105 & 1.105 & 116.025 & 0.116 & 4 \\
109 & 1.105 & 120.445 & 0.133 & 4 \\
114 & 1.105& 125.970 & 0.150 & 4 \\
116 & 1.105 & 128.180 & 0.166 & 4 \\
77 & N/A & 65.848 & 0.033 & 5 \\
100 & N/A & 84.619 & 0.066 & 5 \\
120 & N/A & 101.812 & 0.099 & 5 \\
135 & N/A & 114.802 & 0.133 & 5 \\
145 & N/A & 126.138 & 0.166 & 5 \\
157 & N/A & 134.031 & 0.199 & 5 \\
168 & N/A & 143.734 & 0.233 & 5 \\
115 & 0.565 & 64.975& 0.033 & 6 \\
155 & 0.565 & 87.575 & 0.066 & 6 \\
190 & 0.565 & 107.350 & 0.099 & 6 \\
205 & 0.565 & 115.825 & 0.133 & 6 \\
224 & 0.565 & 126.560 & 0.166 & 6 \\
233 & 0.565 & 131.645 & 0.199 & 6 \\
249 & 0.565 & 140.685 & 0.233 & 6 \\

\noalign{\smallskip}\hline
\end{tabular}
\caption{Fireball evolution in the first 230 ms. The first column shows the radius of the fireball in pixels; the second column shows the pixel scale calculated for each video. Column 3 shows the radius in meters. Column 4 shows the time from explosion for each frame. Column 5 shows the videos labels. For V5 refer to appendix.}
\label{tab:2} 
\end{table}
\begin{figure*}[ht]\centering
    \includegraphics[width=0.8\textwidth]{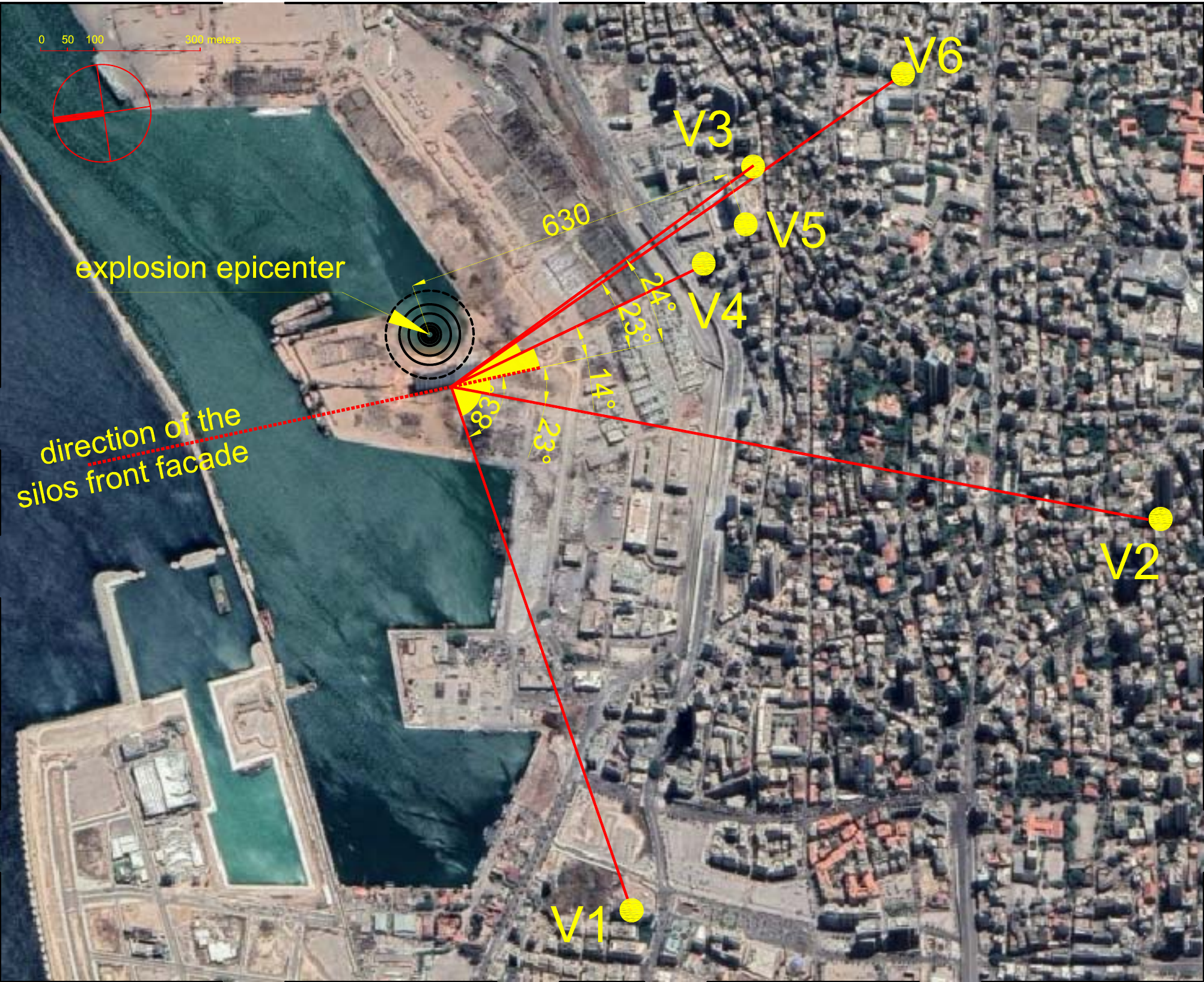}
    \caption{A Google earth map of Beirut showing the location of the 6 videos used in the current report. For each video an incident line of site is determined along the long facade of the grain silos building. Video 5 is calibrated using another technique since this video does not show the silos building in the frames (see appendix). The explosion center is marked with black circles.}
    \label{fig:2}       
\end{figure*}
%%%%%%%%%%%%%%%%%%%
%%%%%%%%%%%%%%%%%%%%%%%%%%%%%%

%
\subsection{Data}
\label{subsec:3}
The data used in the current analysis consist of six different videos taken by smartphones from various locations overlooking the explosion site. The videos used are located on the general map in fig \ref{fig:2} and are listed in table \ref{tab:1}. The videos are taken at different frame rates (FPS) which are also shown in table \ref{tab:1}. The time separation between extracted still images is thus constrained by this limitation.
%%%%%%%%%
%%%%%%%%%%%%%%
\subsection{Measuring the time evolution of the fireball}
\label{subsec:4}
What we exactly need in this study is to trace the evolution of the fireball radius $R_t$ as a function of time  $t_0$ which is considered to be the time zero of the explosion.

\begin{figure}[ht]
  \includegraphics[width=0.48\textwidth]{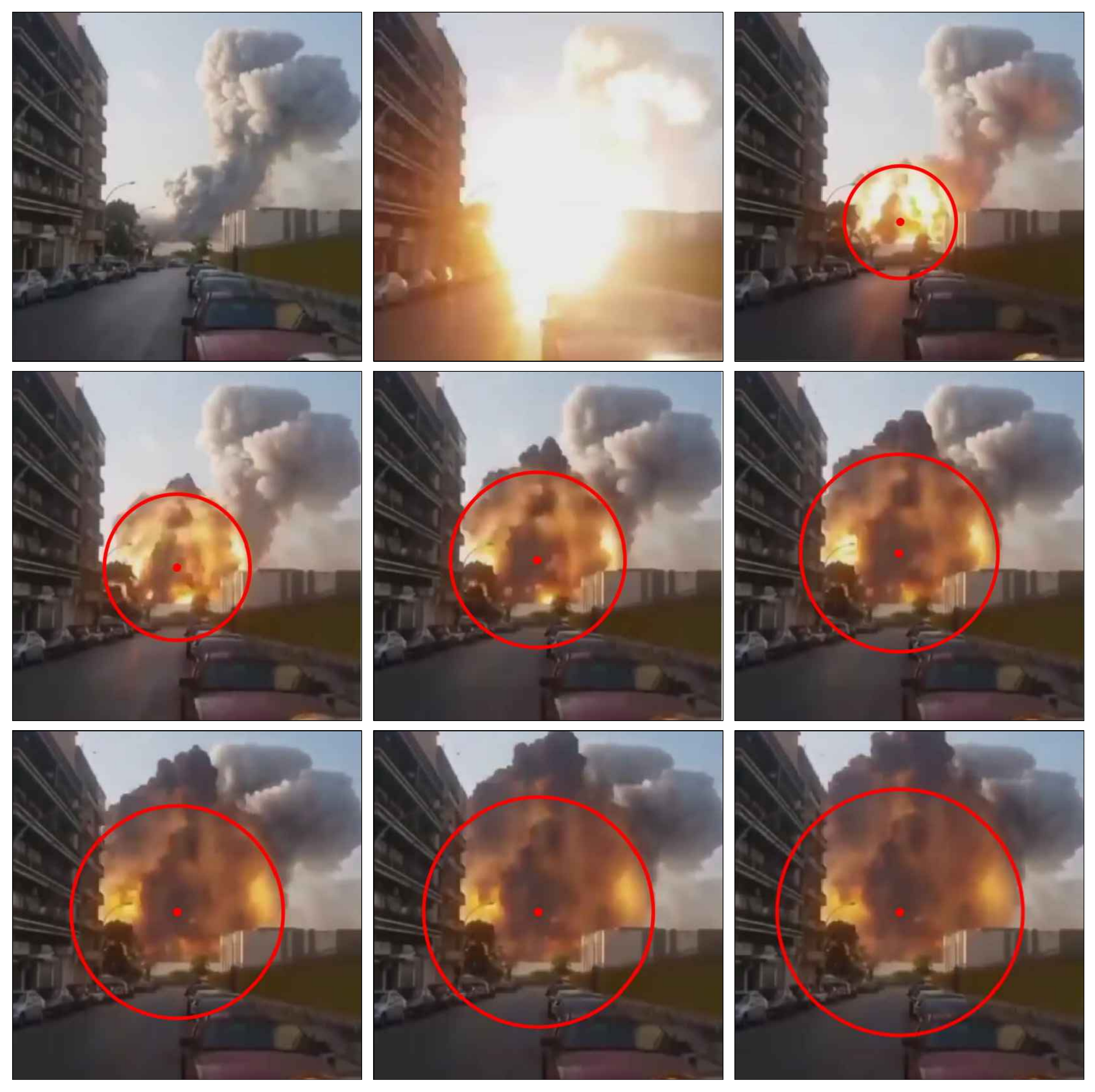}
\caption{Video 5 taken from a distance of 630 meters from the explosion. 9 frames separated by 16.66 ms showing the fireball along with the circle fit. The detonation is assumed to have happened anytime between the first and the second frame. For pixel calibration method see appendix.}
\label{fig:3}       
\end{figure}

 We use the building of the grain silos to calibrate the pixel scale of the videos by defining the location of each footage and defining the incident line of sight angle taken with respect to the grain silos building long facade. We use its accurate \CA{as-built drawings} to de-project the \CA{width, the length and the height} whenever these three are visible. We determine accordingly a pixel scale $\theta$ from both the width and the length $\theta_{\mathrm{[L+l]}} $ and separately from the height $\theta_{\mathrm{[h]}}$, we then calculate an average of these two $\theta_{\mathrm{[mean]}}$ in which a pixel corresponds to a physical measurement in meters.\\

\CA{Here we highlight the precision of our scaling method and the differences with  \textcite{diaz2020explosion}:}
\begin{enumerate}
\item \CA{ We use not only the length of the silos, rather our method combines the de-projected length, width and height. This method reduces the systematic errors associated with calibration. We take a mean value between the three measurements. Our results show excellent convergence to within not more than few meters.}
\item \CA{We use the accurate as built drawings of the silos building and we do not use google map images to calibrate its length. The accurate drawings have been provided through private communication with the local port authorities.}
\item \CA{The building is not a pure rectangle; it has semi cylinders projected from the sides. These cylinders have to be taken accurately when performing a detailed projection. We accurately overlay the rotated building as per our incident angles over the images and make sure that all projected cylinders and edges are aligned with the rotated drawings. Thus we limit the propagation of errors associated with this process. This is clearly seen in fig \ref{fig:4} and in the appendix. }

\item \CA{Different than \textcite{diaz2020explosion}, we trace the fire ball as representative of the shock wave radius only in the first few frames where its boundaries are clearly defined with a reasonable error margin up to about 170 ms after the explosion. The Fireball cannot track the shock wave at all times, therefore, we provide scaling arguments that the epoch within which we use the fire ball as representative of the shock wave and consequently their detachment time are valid (refer to section \ref{subsec:7}). Thus we limit our data points to within this limitation.}
\end{enumerate}
\par We measure the total length of the building projection in pixels as measured on the CCD. This includes the side elevation length $l=30.5$ m (including the half cylinder projection as it defines the border of the projected visible width), and the front elevation length $L=152$ m. The total projected length as measured on the camera plane is thus given by $L\sin{\alpha} + l \cos{\alpha}$ where $\alpha$ is the incident angle of sight. A pixel scale $\theta_{\mathrm{[L+l]}}$ is then given by $\frac{L\sin{\alpha} + l \cos{\alpha}}{L_{\mathrm{total[pixels]}}} $. \CA{Wherever the ground is visible, we use the height of the silos as an additional scaling parameter and we compute the height scaling parameter $\theta_{\mathrm{[h]}}$, we then take the average between these two $\theta_{\mathrm{[mean]}}$}. This is shown in Fig. \ref{fig:4}, and in table \ref{tab:3} and \ref{tab:2}. The physical value $R_\mathrm{m}$ in meters, will thus be given by:

\begin{equation} 
\label{eq7}
    R_m = \theta_{\mathrm{[mean]}} \times R_{\mathrm{[px]}}.
\end{equation}

Here $R_\mathrm{m}$ is the radius of the fireball in meters and $R_{\mathrm{px}}$ is the measured radius in pixels. The values of $\theta$ are shown in table \ref{tab:2}.

\begin{figure*}[ht]
 
  \includegraphics[width=1\textwidth]{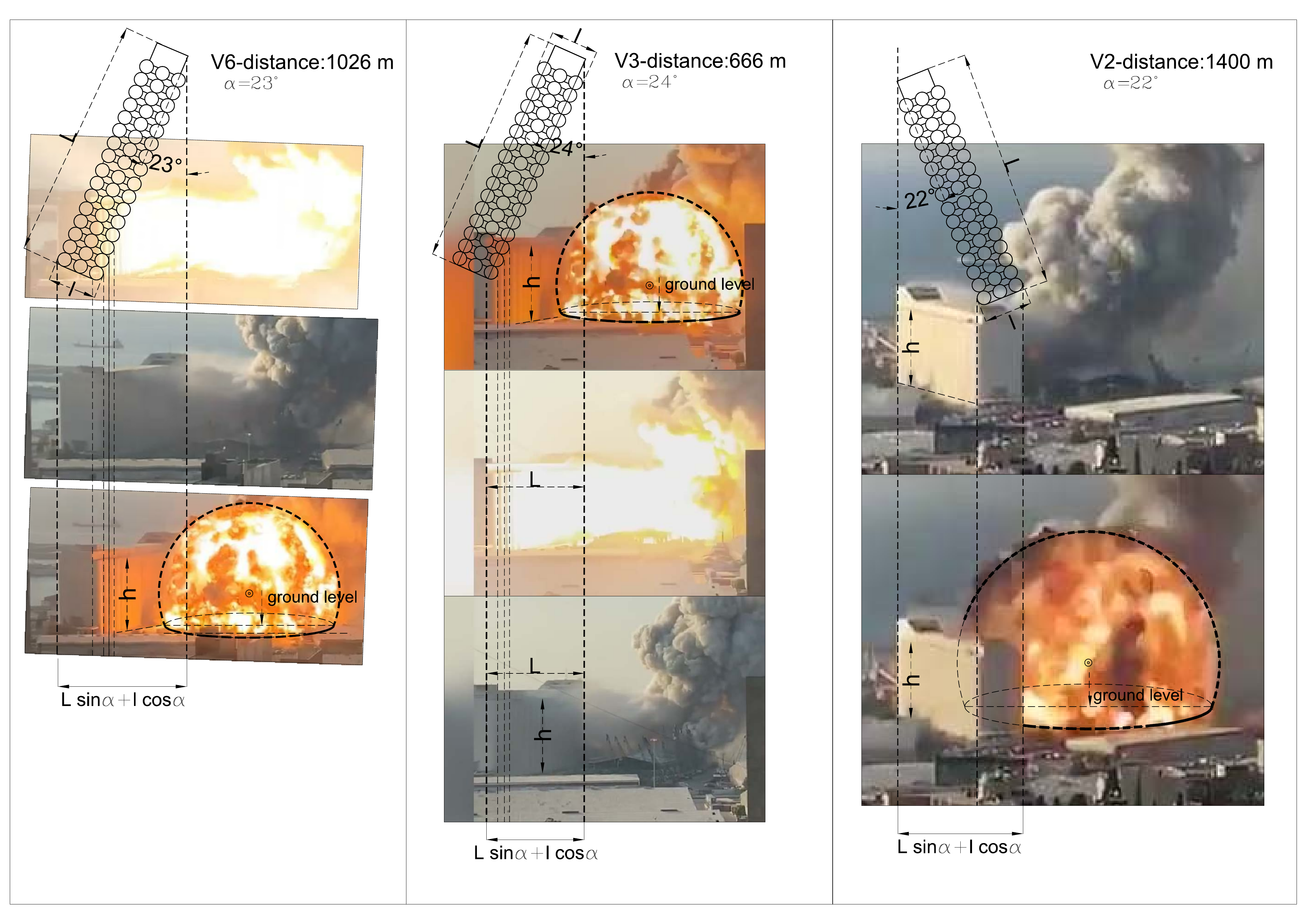}
\caption{For each video the incident angle of sight towards the silos long facade is shown along with the actual picture of the silos and the de-projection lines. Drawings are provided by the \textcite{aca}.}
     \label{fig:4} 
\end{figure*}

\begin{table*}[ht]\centering

\begin{tabular}{lllllllllll}
\hline\noalign{\smallskip}
Video & $L$ & $l$ & $\alpha$ & $L \sin\alpha+l\cos\alpha$ & $L+l$ & h & h & $\theta_{[L+l]}$ & $\theta_{[h]}$ & $\theta_{\mathrm{[mean]}}$\\
 &[meter] & [meter] & [$^{\circ}$] & [meter] & [pixel]& [meter] & [pixel] & $\frac{\mathrm{meter}}{\mathrm{pixel}}$ & $\frac{\mathrm{meter}}{\mathrm{pixel}}$ & $\frac{\mathrm{meter}}{\mathrm{pixel}}$\\
  
\noalign{\smallskip}\hline\noalign{\smallskip}
1& 152 & 30.5 & 83 & 154.592 & 109 & 48 & 36 & 1.413 & 1.333 & 1.373 \\
2& 152 & 30.5 & 22 & 85.194 & 73 & 48 & 43 & 1.167 & 1.0116 & 1.141\\
3& 152 &  N/A   & 24& 61.794 & 182 & 48 & 138 & 0.339 & 0.347 & 0.343 \\
4& 152 & 30.5 & 14 & 66.348 & 60 & N/A & N/A & 1.124 &N/A  & 1.105 \\
6& 152 & 30.5 & 23 & 87.440 & 156 & 48 & 91 & 0.602 & 0.527 & 0.565 \\
\noalign{\smallskip}\hline

\end{tabular}
\caption{Calibration parameters.The first column shows the video label. The second and the third columns show the length and the width of the silos in meters, the third shows the incident angle, the fourth shows the total projected length against the line of sight, the fifth shows the total projected length in pixels as measured on the CCD, the seventh and the eight show the height of the silos in meters and pixels respectively. The last three columns show the pixel scale for the values derived respectively  from [$\mathrm{L+l}$], $\theta_{\mathrm{[L+l]}}$, from[h], $\theta_{\mathrm{[h]}}$ and the final mean value $\theta_{\mathrm{[mean]}}$ used for the final calibrations.
The width $l$ for V3 is not visible in the footage and the height for V4 could not be measured because the video does not show a clear view to the ground. V5 is calibrated using another method (refer to the appendix).}
\label{tab:3} 
\end{table*}

\par This procedure does not apply to Video 5 in which the grains silos building is not visible. To calibrate this video, we determine the angular field of view in which a pixel corresponds to an angle resolution. Using this angular pixel resolution, we can calculate the dimensions of the fireball knowing its distance (630 meters) using basic trigonometry. This is described in the appendix. Surprisingly, we find comparable numbers with different videos taken from completely different locations and using different scaling methods. 
\par The separation between consecutive frames changes with the videos. For most of them (video 1-2-3-4-6), the rate is 30 FPS corresponding to time intervals 33.33 ms. For video 5, the rate is 60 FPS corresponding to time intervals of 16.66 ms. 
\par For each frame, we measure the size of the fireball by manually fitting a circle to the luminous edge and converting the pixels to physical measurements in meters using the pixel scale $\theta_{\mathrm{[mean]}}$ derived earlier. \CA {Measurements are executed to the largest visible width of the projected visible sphere.(refer to fig \ref{fig:4})}\\
It is important to note that some of the videos (namely 1,2 and 4) were taken by a shaking hand. In that case, defining the center of the explosion is executed taking a fixed reference from the picture for which the coordinates of the center are corrected for each and every frame. We identify the time of the explosion $t_0$ by visually checking the frame at which the first bright light of the explosion is seen. Thus, $t_0$ can be taken anytime between this frame and the previous frame. We consider an error on determining $t_0$ of 16 ms. (We refer the reader to section \ref{subsec:5} for a complete error analysis review).\\
Using this procedure we can build the time evolution of $R_{{t}}$ with respect to $t_0$.\\
The luminous sphere is bound by a pseudo-sharp edge in the first 100 ms, as this edge starts becoming less sharply defined at later stages, we do not extend our measurements further. Our values are shown in table \ref{tab:2}.

%%%%%%%%%%%%%%%%
%%%%%%%%%%%%%%%%%

%
\subsection{Error analysis}
\label{subsec:5}
Two main uncertainties can cause the error on the estimation done in the current study, namely the error on measuring the radius $R$ and the error in assuming the time $t_0$ of the detonation.\\
We are measuring pixels in frames extracted form videos taken by handheld smart phones of limited frame rates cameras.
\par
\begin{figure}[ht]
  \includegraphics[width=0.5\textwidth]{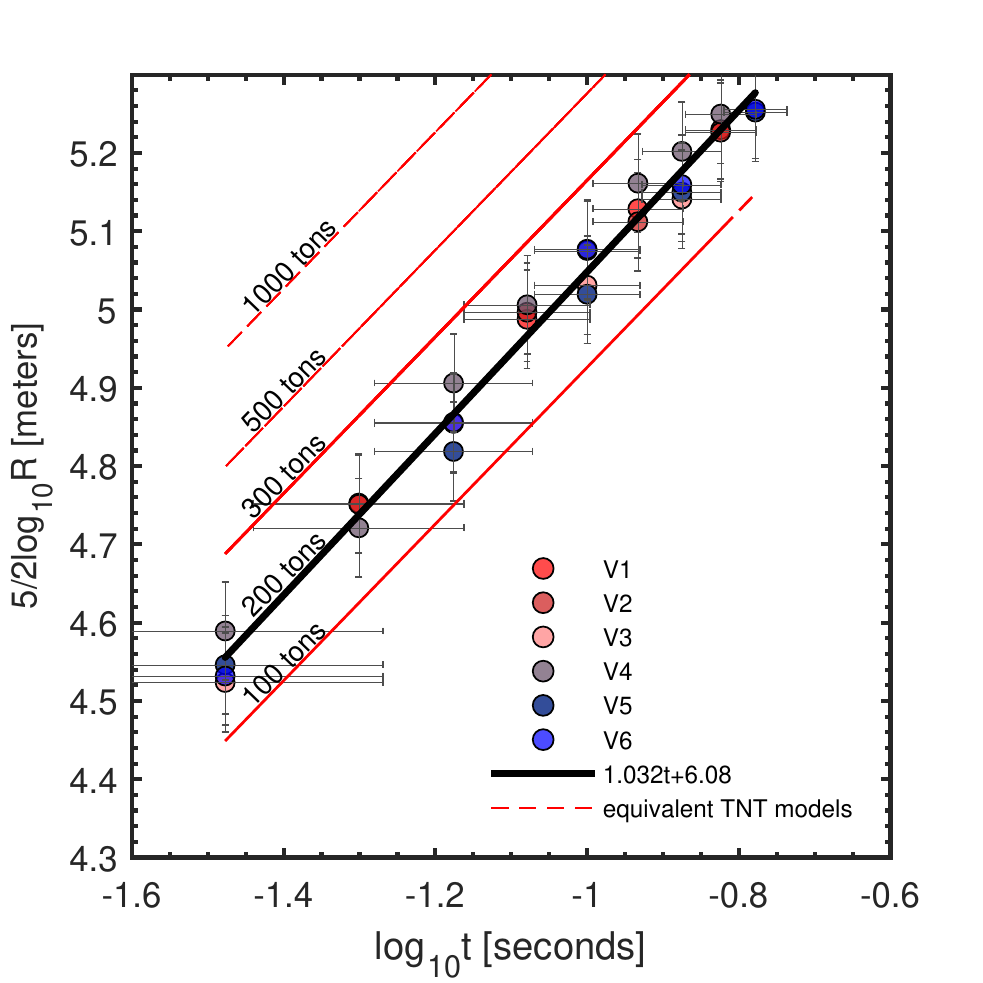}
\caption{Fireball radius evolution with time. $\frac{5}{2}\log_{10}R$ on the $Y$-axis and $\log_{10}t$ on the $X$-axis. The dots represent the fireball radius taken from different videos for data points taken at a distance less than about 130 m at an epoch less than about 170 ms  where the fireball is assumed to trace the shock wave. The best linear fit has the form $Y=ax+b$ where $a$=$1.024$[0.955-1.092] and $b = 6.052$[5.977-6.127]. The red lines represent the equivalent TNT yields from surface blast explosions: i.e. corrected by a factor of 1.8 to include reflection from the ground.}
\label{fig:5}       
\end{figure}

\begin{figure}[ht]
  \includegraphics[width=0.5\textwidth]{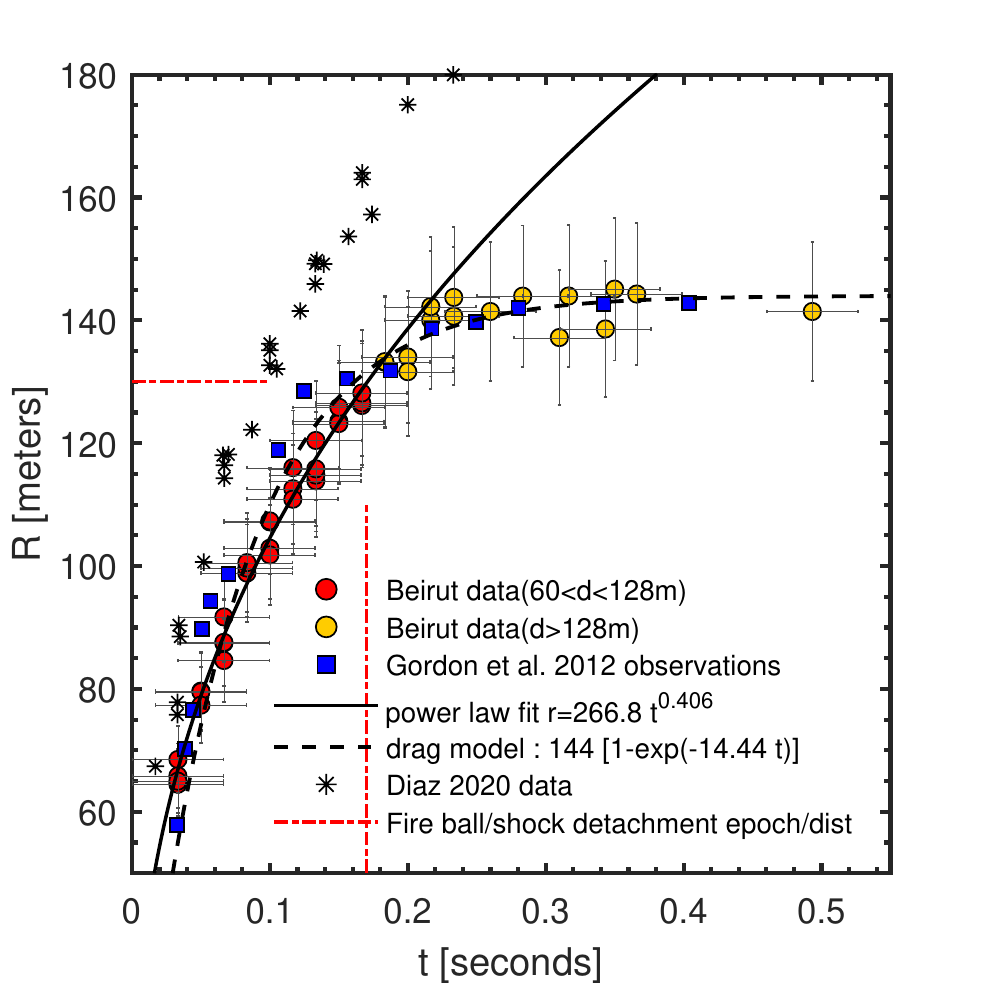}
\caption{The red dots are the fireball measurements used in yielding the TNT equivalence of the explosion, taken at distance less than about 130 m at times less than about 170 ms. A power law of exponent 0.406 which is represented via the black line fits the data at this epoch and is consistent with a Sedov-Taylor model with spherical geometry and instantaneous energy release. At this epoch the fireball measurements are assumed to measure the shock wave evolution. The orange dots are the fireball measurements taken at later epochs: i.e. where the fireball and the shock wave have already detached, until the luminous fireball has reached its final radius. The dashed curve represents a drag theoretical model describing the evolution of the fireball until reaching the stopping distance at about 140 m. The straight dashed red lines represent the epoch and the distance where the shock wave is assumed to detach from the fireball. The blue squares represent the fireball evolution from \textcite{gordon2013fireball} scaled to the TNT equivalence of Beirut explosion. Small asterisks show the \textcite{diaz2020explosion} data for comparison.}
\label{fig:6}       
\end{figure}

 These errors can be divided into three main categories.
\begin{enumerate}
    \item Errors in determining the pixel scale: this is mainly due to uncertainties in defining the precise location of the camera within at least $1^{\circ}$ and errors in defining the boundaries of the silos border due to noise in the frames and the low resolution for some of the videos. The combined effect is assumed to cause an uncertainty in the pixel scale determination of $\frac{\delta \theta}{\theta}\approx 5\%$.
    \item Errors in fitting the circles and measuring the pixel size of the fireball: this is especially at later times when the edge of the fireball becomes less sharply defined. We estimate this uncertainty to about 4 pixels as an average from all measurements. This leads to an average value taken from all measurements $\frac{\delta R_{\mathrm{pixels}}}{R_{\mathrm{pixels}}}=3\%$. Using eq \ref{eq5} we can write: 
\begin{equation}
\label{eq8}
    \frac{\delta R_\mathrm{m}}{R_\mathrm{m}} = \sqrt{\left(\frac{\delta \theta}{\theta}\right)^2 + \left(\frac{\delta R_{\mathrm{pixels}}}{R_{\mathrm{pixels}}}\right)^2}.
\end{equation}
Here $\delta R_m$ is the error on the radius of the fireball in meters and $\frac{\delta\theta}{\theta}$ is the error on the pixel scale. Taking the average relative pixel scale $\frac{\delta\theta}{\theta}$ of 5\% and an average $\frac{\delta R_{\mathrm{pixels}}}{R_{\mathrm{pixels}}}$ of 3\% leads to a relative error $\frac{\delta R_{\mathrm{m}}}{R_{\mathrm{m}}} = 0.058$.\\
Thus the propagation of error on the value $\frac{5}{2}\log_{10}R$ is given by differentiating with respect to $R$ leading to \begin{equation} 
\label{eq9}
\delta\left[\frac{5}{2}\log_{10}R_\mathrm{m}\right] = \frac{\delta R_\mathrm{m}}{R_\mathrm{m}}  \frac{5}{2\log_\mathrm{n}{10}} \approx 0.063.
\end{equation}
    \item Error on the assumption of time $t_0$ of the detonation: by checking the frames, it can be seen that the detonation causes a sudden bright light seen in the frames sequence, thus, the detonation is assumed to happen anytime between these two consecutive frames (after the frame without intense light and before the frame with the intense light) which can be seen in the first two frames of Figures \ref{fig:3}-\ref{fig:7}. We consider an average error on $t_0$ of $\delta t \approx \pm 17$ ms.\\
    The error on the term $\log_{10}t$ is given by \begin{equation} 
    \label{eq10}
        \delta \log_{10}t = \frac{1}{\log_\mathrm{n}{10}} \frac{\delta t}{t} = 0.434 \frac{\delta t}{t}.
\end{equation}

\end{enumerate}
Taking the combined effect of these errors will cause a very large error propagation on the value of the energy $E$ if this will be computed only using equation \ref{eq1} with only few measurements of $R$ and $t$. Our 39 measurements and the linear fitting will reduce the effect of these errors and will limit the error to only the one on the fitting parameters, namely the slope $a$ and the intercept $b$ of equation \ref{eq2}.  

\section{Results and Observations}
\label{sec:3}
A clear sharply defined hemispherical fireball was visible in approximately the first 100 ms after the detonation. The boundaries of its expanding surface became less sharply pronounced \CA{as it expanded}. Later, its boundary rapidly faded until it was not clearly defined after about  170 ms post-detonation. This is partly due to the rapid cooling caused by the rapid decrease of over-pressures, and partly due to the contamination of the atmosphere with the existing dark smokes caused by the first fire and also to the turbulent instabilities created at the boundary of the hot sphere with the surrounding atmosphere. Nevertheless, the boundary of this sphere can be reasonably traced within an acceptable error estimate (few pixels). This is possible within the first few frames where the fireball boundaries are reasonably defined as it expands all the way until the expansion slows down and seems to halt.
\CA{We limit our data for the energy estimation to an epoch less than 170 ms, at a distance less than 130 meters.}

After this, the fireball and the shock wave depart and the former does not trace the latter any longer (we refer the reader to section \ref{subsec:6} for a complete discussion). At this stage, when the temperatures drop, the shock wave causes the formation of the vapour cloud, a.k.a Wilson cloud, clearly spotted in the videos rapidly growing, leading a massive damage front ahead of its boundary. The formation of this cloud is caused by the negative phase following the passage of the sudden increase in pressure, and the adiabatic cooling causing the atmospheric water vapour to condensate and create the white appearance. It is important to mention that at these epochs the boundary of the shock wave is not traced by the boundary of the Wilson cloud due to the delay between the positive pressure and the negative pressure phase. Condensation will only happen during the negative pressure phase. Therefore, using the vapour cloud to trace the shock wave cannot lead to significant results. (Tracking the evolution of the condensation cloud is the subject of future work.)
\par \CA{We use 30 measurements of radius $R_{t}$ at distances less than 130 meters and time $t$ at less than 170 ms, and we fit a power low to the data. Our best fit is a power law of the form $r=at^b$, where $a=266.8$ and $b= 0.406$. Assuming a spherical geometry (as clearly seen from the fireball shape), we conclude $n=3$ which leaves us with $s\approx 0$ (ref to Eq. \ref{eq6}). This is very close to the assumption of instantaneous energy release. Therefore, we conclude that the Taylor model is valid within this range (we refer the reader to sec \ref{subsec:6} for further discussions)}.
\par We now fit the values $\frac{5}{2}\log_{10}R$ on the $Y$-axis and the value $\log_{10}t$ on the $X$-axis. The data is consistent with the theoretical prediction, this fact is remarkable knowing the cumulative error margin due to the quality of the data and the limitations of our procedure. 
Our best fit to the data is a line of constant gradient of the form $\frac{5}{2}\log_{10}R = a\log{t} + b$ where \CA{$a = 1.024$ within an error range of [0.955-1.092] and $b = 6.052$ within an error margin of [5.977-6.127]. Taking $K = 0.856$ as given by \textcite{taylor1950formationI} for a diatomic gas where $\gamma = 1.4$ and $\rho_0 = 1.23\ \ \mathrm{kg.m}^{-3}$ and by using Eq. \ref{eq3}, we find the value of $E$ to be $1.53\pm 0.6\times 10^{12}$ Joules or the equivalent of $365\pm 143$ tons TNT. Here we use a conversion factor between Joules and TNT equivalent of $4.184\times 10^9 \frac{\mathrm{Joules}}{\mathrm{TNT}}$ \parencite{karlos2013calculation}.} 
\CA{However, since this is a surface explosion, the reflected shock from the ground: i.e. the fraction of its energy that has not been used into cratering, will immediately catch up with the expanding shock in air, and produces a stronger Mach wave, creating an enhancement of the over-pressures, and thus will create the appearance of a shock wave that is larger than one originating from a free air burst with no reflected surfaces \parencite{kingery1984airblast,cormie2019blast,baker1973explosions}. \citeauthor{kingery1984airblast} \cite{kingery1984airblast} consider that 10 \% of the energy is absorbed by the ground while the remaining is reflected and thus a factor of 1.8 has to be used.}
This will yield an energy of $E\approx 0.85\pm 0.3\times10^{12}$ J or an equivalent of $ 200 \pm 80$ tons TNT. 
\par \CA{The ammonium nitrate conversion to TNT is variously reported in the literature (56 \% \parencite{ANequiv2merrifield1991comparison} 38\% \parencite{karlos2013calculation} and recently 42\% \parencite{ANequiv3braithwaite2020ammonium}). Even if we use the lowest conversion value of \textcite{karlos2013calculation}, we estimate that the ammonium nitrate equivalence of Beirut explosion is $526\pm 210$ tons. This is much less than the quantity of ammonium nitrate claimed by the local official authorities records of 2750 tons originally stored at the port. This can be explained by two scenarios: either a significant fraction of the stored quantity was consumed by a deflagration before the detonation occurred and thus had limited input in the blast energy, or the missing quantity was not physically available at the time of the explosion. The current study cannot confirm one of these two scenarios.}
Here it is important to mention that this approach does not consider the amount of energy that has been radiated as heat in the form of electromagnetic radiation, however, these effects are considered insignificant in a chemical explosion \parencite{bethe1958blast}, contrary to a nuclear explosion. Moreover, part of this energy has been used in demolishing the steel structure hangar in which the explosives where kept, which is also assumed to be insignificant.
\CA{ In general, the errors accumulated because of these assumptions will still be contained within the error margin within which we state our estimation.} 

\subsection{The applicability of the Sedov-Taylor model} 
\label{subsec:6}
\par \CA{It is well known that Taylor model assumes a point source solution in which the mass of the explosives is insignificant and the release of energy is assumed to be instantaneous. Although this approximation is valid for a nuclear explosion, it is not valid for a chemical explosion because these assumptions do not hold. However, using experimental data and comparing it to theoretical work, \textcite{taylor1950formationI} has provided a range where a chemical explosion behaves similar to a nuclear explosion and the two can be comparable.}
This range is provided in \textcite[(see page 170-173- Fig. 5)]{taylor1950formationI}
and is given as a function of the scaled distance $\log{\left[R E^{-\frac{1}{3}}\right]}$. Here $E$ is the energy yielded from the explosion given in ergs, and $R$ is the distance in cm. Taylor proposes that the window of comparison lies between values of $\log{\left[R E^{-\frac{1}{3}}\right]} = -2.3$ to about $-2.7$. 
\par
\begin{figure}[ht]
  \includegraphics[width=0.45\textwidth]{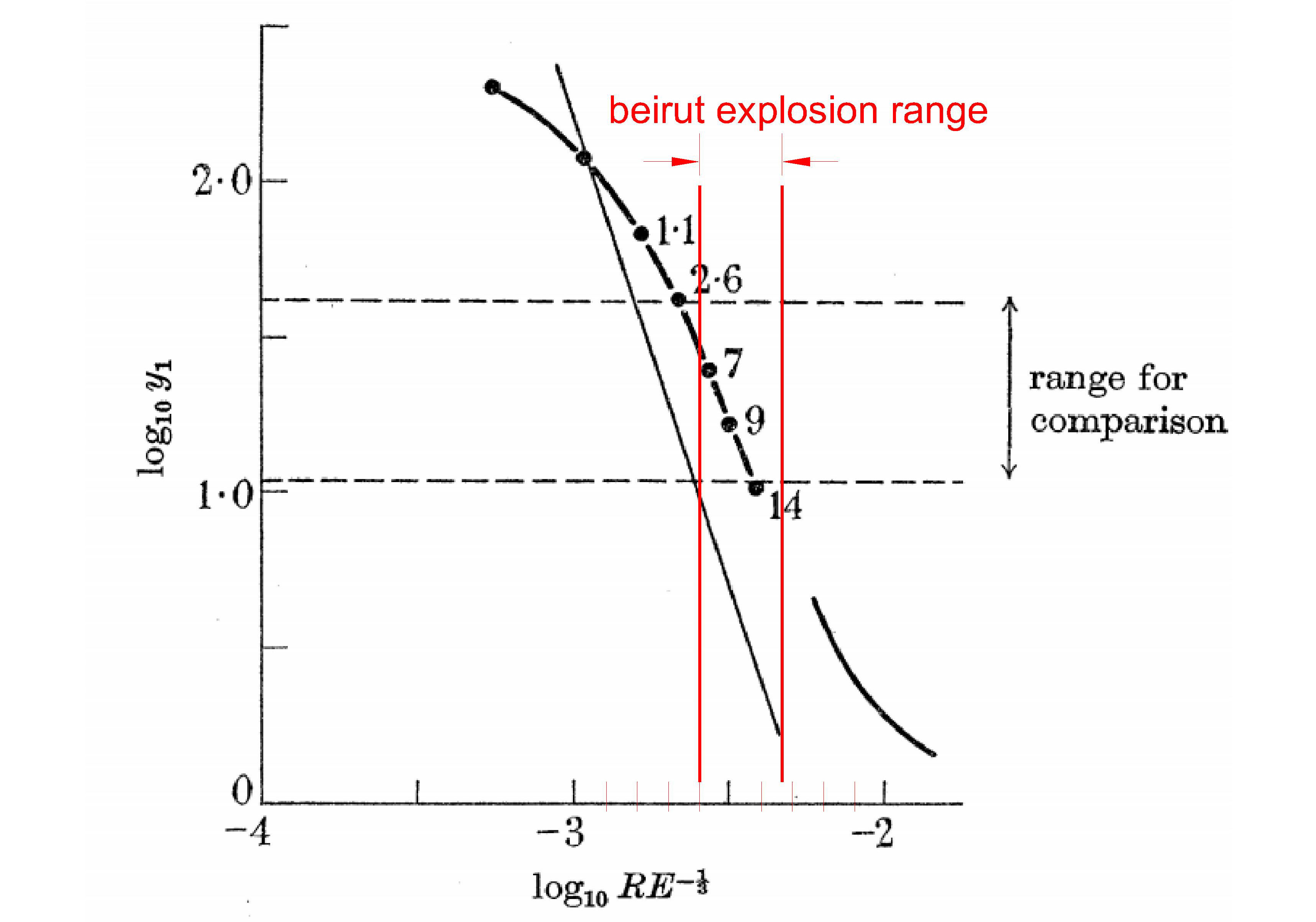}
\caption{Original data from\textcite{taylor1950formationII}[see fig5]. The continuous straight line shows the blast pressures on the Y axis derived from a theoretical Taylor model, the thick line with the dots are data taken from observations for a chemical explosion with the same release of energy. The X axis shows the $\log_{10}$ of the scaled distance $\log{\left[R E^{-\frac{1}{3}}\right]}$. The Vertical red lines show the boundaries of Beirut data between distances of 60 and 120 m scaled to our TNT equivalent yield.}
\label{fig:7}       
\end{figure}

  We perform the necessary units conversion and compare our results for the range of our distances (60-130 m), we find values of $\log{\left[R E^{-\frac{1}{3}}\right]}=-2.6$ to $-2.3$ for distances between 60 and 130 m. This falls comfortably within the comparison range. This is shown in fig \ref{fig:7}. 
It is worth stressing that our data show a remarkable consistency. Points taken in a distance range of $~60-130$ meters still follow the trend as expected from theory. Note that, as previously stated, this data are severely limited (taken from 6 different videos at 6 different locations using different approximate scaling and limited by the resolution).\\

\subsection{The fireball kinematics and stopping distance compared with experimental observations}
\label{subsec:7}
It can be seen in Fig. \ref{fig:6} that the radius of the fireball decelerates until it stabilizes at some distance $R_\mathrm{m} \approx 140-145$ m after about $200$ ms. The shock wave at this epoch \CA{has already detached} from the fireball and moves ahead of it. This has been observed before \parencite[see][]{taylor1950formationII, mack1946trinityphotos, szasz1984daytrinitytest, gordon2013fireball}. At this epoch the shock wave is driven by the remaining energy that has not been radiated as heat or consumed during the expansion of the heated air. The epoch during which the shock wave can be traced by the fireball is thus limited.\\

\par We use a drag model \parencite{gilev2006interaction,gordon2013fireball} to \CA{describe} the kinematics of the fireball. This is given by:  
\begin{equation} 
R_\mathrm{f}(t) = R_\mathrm{m}[1-\mathrm{exp}(-\kappa t)].
\end{equation}
Here $R_\mathrm{f}(t)$ is the radius of the fireball, $R_\mathrm{m}$ is the stopping distance, a distance at which the fireball expansion becomes asymptotic, and $\kappa$ is the drag coefficient. The drag model with $\kappa = 14.44$ and $R_\mathrm{m} \approx 144$ m is in good agreement with the observed evolution of the fireball.
This can be clearly seen in fig \ref{fig:6}. One can notice that the drag model curve follows the observed points within the range of error. The drag model significantly departs from the Sedov-Taylor model at about 100-130 m, a point after which the assumption of the fireball tracing the shock wave does not hold any further for the evaluation of the energy.

\CA{  \textcite{gordon2013fireball} measure separately the evolution of both the fireball and the shock wave using high speed photographs from several charge loads sufficiently elevated from the ground. On Fig. \ref{fig:6}, we plot our data compared with the observations of \citeauthor{gordon2013fireball} scaled using Hopkinson's cube root scaling $\frac{R_1}{R_2}=(\frac{W_1}{W_2})^\frac{1}{3}$. It can be seen that the evolution of the two fireballs are very similar. For a charge load of about 20 kg TNT equivalence, the fireball reaches a stopping distance at about 5.5 m, this corresponds to a scaled distance $Z=\frac{R_\mathrm{m}}{W^{\frac{1}{3}}}=2.02 \frac{\mathrm{m}}{\mathrm{kg}^{\frac{1}{3}}}$. Taking our TNT equivalent yield and using the same scaled distance we find a stopping distance of about 145 m. This is in excellent agreement with our results and provides an additional check on the validity of our findings.}\\
\CA{Furthermore, \textcite{gordon2013fireball} find that the distance at which the sock wave detaches from the fireball is about 5 m for a charge load of 20 kg, this corresponds to a scaled distance $Z\approx 1.84\frac{\mathrm{m}}{\mathrm{kg}^{\frac{1}{3}}}$. Using this scaled distance and our derived TNT equivalent yield, we find a detachment distance of about 130 m also in perfect agreement with the range over which we used the fireball to trace the shock wave.}\\

\CA{It is worth to mention that the fireball hemisphere does not stay centered on the ground, and this is clearly seen in the frames (see fig\ref{fig:4}). Part of this is due to buoyancy and part is due to the reaction from the ground. This has been observed before \parencite{gordon2013fireball, taylor1950formationII, mack1946trinityphotos} . \citeauthor{gordon2013fireball} \cite{gordon2013fireball} measure the lift of the fireball. In the first 3 ms the fireball has already reached a height of about 4.5 m for a charge weight of 20 kg TNT equivalence. Using Hopkinson's cube root scaling for our TNT equivalent value this corresponds to a height of 115 m at around 100 ms. From observing the height of our fireball by comparing it to the silos building top level in the first 3 frames, we find a value of about $120\pm10$ m in the first 100 ms which is consistent with \textcite{gordon2013fireball} observations. Unfortunately the height of the fireball was obscured by the dark smokes and could not be measured at later times.}\\

\par The follow-up of the shock wave beyond this epoch is out of the scope of the current work. However it is worth noting that such a follow-up is stringently limited and tracing it with these videos is challenging for the following reasons:
\begin{itemize}
\item[$\bullet$] Tracing the shock wave using the vapour cloud is not consistent because of the delay in the formation of the vapour after the passage of the positive phase of the over-pressure. This time delay makes the actual location of the pressure front ahead of the vapour cloud. Additionally, pixel scaling set with reference to the silos building holds only in the vicinity of the explosion center. However, this cannot be accurate for extended distances where the dimensional pixel scaling does not hold anymore.

\item[$\bullet$] Tracing the shock wave using time of arrival of the shock from frame by frame analysis of the videos may be affected by the urban pattern. In fact, the interaction of the shock wave with the dense urban structures causes complex interactions and diffraction patterns, making the assumption of an isotropical spherical flow of the shock wave inaccurate \parencite{shockdiff1,shockdiff2,shockdiff3}. The interaction of the shock wave with the existing grains silos, for example, has created a visible dark spot within the vapour cloud, signaling the absence of strong over-pressures and therefore the absence of the shock wave in this particular location. Later this spot has disappeared due to interference of the reflected wave from the ground and the edges of the building. This can be clearly seen in fig \ref{fig:12} in the last panel.
Furthermore, at later times, the TNT equivalence cannot be derived using the current method and other relations have to be used (see for example \textcite{rigby2020preliminary}).
\end{itemize}

\section{Comparison with the literature} 
\label{sec:4}
\par \CA{\textcite{dewey2021tnt} calculates the TNT and ANFO equivalences as functions of radial distances from the centre of Beirut explosion. He uses the time of arrival of the shock wave data taken from \textcite{rigby2020preliminary} to derive the velocity of the shock wave. Then using Rankine-Hugoniot relations he derives the hydrostatic overpressures behind the shock and compares the results to experimental data taken from \textcite{deweycompendium}. He shows that Beirut explosion was weaker than a TNT explosion with the same energy close to the center of the explosion. The two blasts became identical at about 500 m (see fig 2 \& 4 of \cite{dewey2021tnt}). At larger distances the Beirut explosion was slightly stronger than that of a TNT. He explains this result by variations in the change of entropy: an originally weaker blast will result in a smaller change in entropy and thus will keep more energy at larger distances.\\
Using this result he showed that the TNT equivalence of the Beirut explosion is an increasing function of distance (see fig 3 of \cite{dewey2021tnt}). It increases from 0.15kt to 0.7 kt TNT between distances of 80 to 1000 meters.} 

\CA{This is in excellent agreement with the results presented in this paper. We find a TNT equivalence for the Beirut explosion of about 200 tons at a distance range between 60 and 130 meters.
Most studies using time of arrival of the shock wave (from audio and visual data)\cite{rigby2020preliminary,stennett2020estimate,pasman2020beirut} report values of 500-700 tons of TNT at distance between 500 and 1000 meters, also in accordance with \textcite{dewey2021tnt}.}

\subsection{The dimensions of the crater}

\CA{The diameter and the depth of the crater are known to correlate with the explosive mass, however, this relation depends on several parameters (nature of the soil and its density, nature of the explosive compound) and is prone to large uncertainties. The literature reports large variations in these relations. } 
 \par
\begin{figure*}[ht]
  \includegraphics[width=1\textwidth]{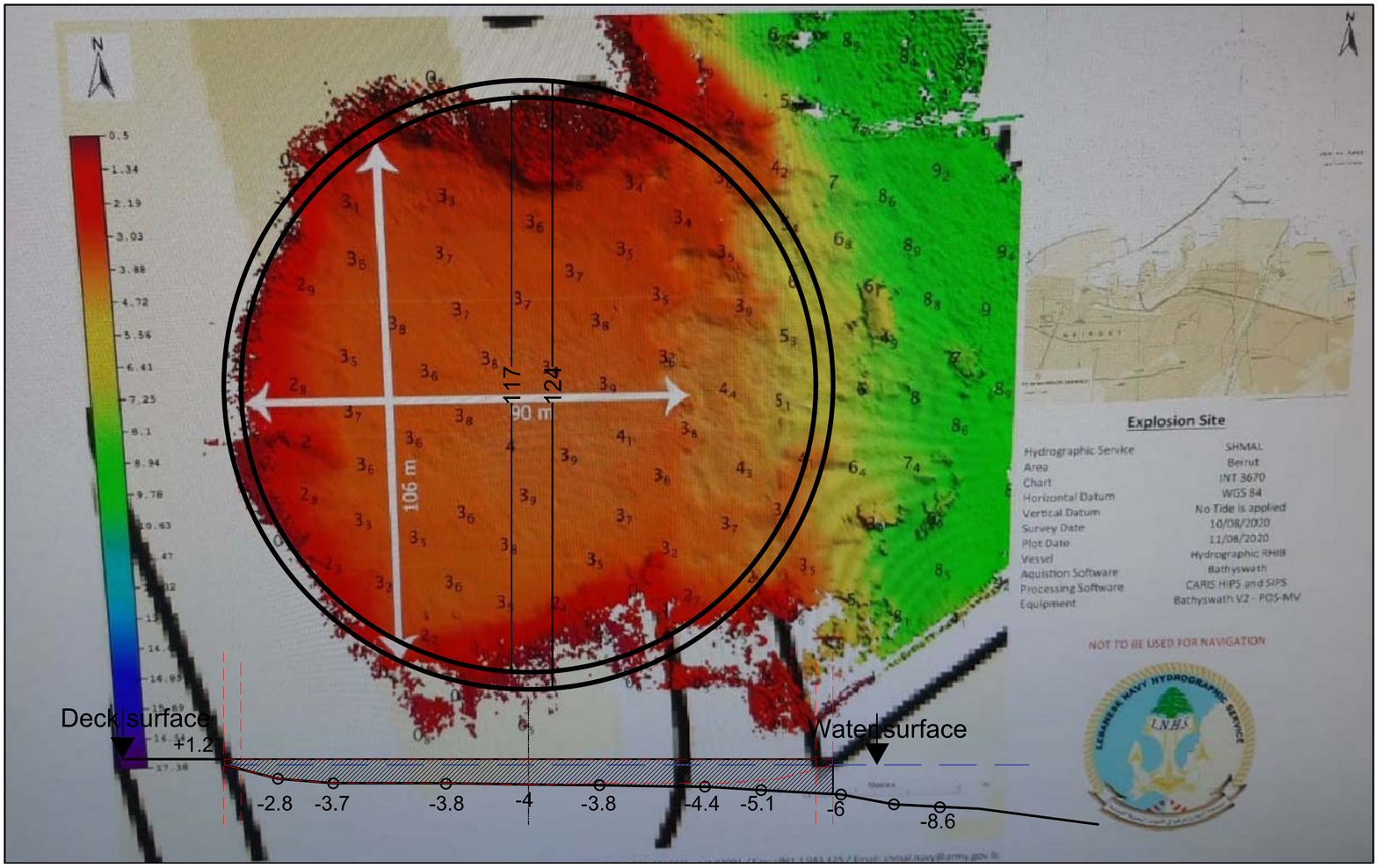}
\caption{Hydrographic survey on the 10/08/2020, 6 days after the explosion. The survey was executed by the Lebanese Navy and communicated to us through a private communication. The deepest point below the water level is 4 meters. Height of the concrete deck is 1.2 meters above water level. A schematic section is shown below. The volume is calculated by extruding the profile along a circular path of 117-124 m diameter.}
\label{fig:8}       
\end{figure*}

 An essay of measuring the energy yield from the crater dimensions has been done by \textcite{pasman2020beirut}. They report a diameter of 124 m and varying values of the depth 13.7, 23 and 43 meters. (The value of 43 meters was reported in the news without any physical proof and the value of 13.7 and 21 meters were scaled from the diameter value). Although scaling relations report a much larger depth/diameter ratio for the craters, Beirut explosion does not seem to follow this trend. In fact, a hydrographic survey was executed on 10/08/2020, 6 days after the event by the Lebanese navy (communicated to us through a private communication and can be seen in fig \ref{fig:8}). The crater shape was an ellipsoid whose major axis was measured to about 117 m and a maximum depth from the surface of the water of only about 4 m.
 \par Here we reuse the same equations as used by these authors, to derive a TNT and AN equivalent mass. The first equation is given by \textcite{cratermannan2013lees} and has the form of $v=0.4W ^{\frac{8}{7}}$ where W is the mass explosives in [lb] and $V$ the volume of the crater in [ft$^3$]. We calculate the volume of the crater by building the sectional profile as shown on the survey and extruding it along a circular path of 117 m and 124 m diameter and noting that the deck is 1.2 m higher than the surface of the water. This yields a total volume range of $42700-48600$ $\mathrm{m}^3$ and therefore an explosive mass of $257-287$ tons TNT. This is in a close range to the upper margin of the results of the current study.
 Another form used by the same authors is given as $V=0.68(W_{\mathrm{AN}})^{0.81}$ where $W_{\mathrm{AN}}$ is the ammonium nitrate mass in kg and $V$ is the volume in $\mathrm{m}^3$. Using the same crater physical volume range we find an ammonium nitrate mass of $840-980$ tons, also in the same order of magnitude of the results of the current study.

\section{ Additional thoughts }
\label{sec:5}

\CA{A question here arises. What is the actual charge load of the Beirut explosion? The TNT equivalence of an event is derived by comparing distances at which physical values such as hydrostatic overpessure, velocity of the shock or impulse are comparable \parencite{dewey2021tnt}. This may not be valid for different explosive compounds. It is sure that all the attempts done in this effort have some degree of uncertainty and each method is associated with a margin of error. The data used for the Beirut explosion is taken from amateur footage and was never designed to conduct an accurate experimental analysis.}

\par \CA{Methods using the time or arrival of the shock wave at large distances suffer from interference with the dense urban structure. Methods using the measurements of the fireball evolution are restricted by the frame rate limits of the smartphones and associated with large margin of errors.}
 \par \CA{Finally, taking the uncertainties and the large error margins of the different methods, the best way would be to model the explosion using the exact urban structure, but then again, this requires an established material model/equation of state of the ammonium nitrate and the actual physical conditions at the time of the explosion. This is beyond the reach and purpose of the current study.
 Much still needs to be learnt.}
 
\begin{figure}[ht]
 \centering
  \includegraphics[width=0.45\textwidth]{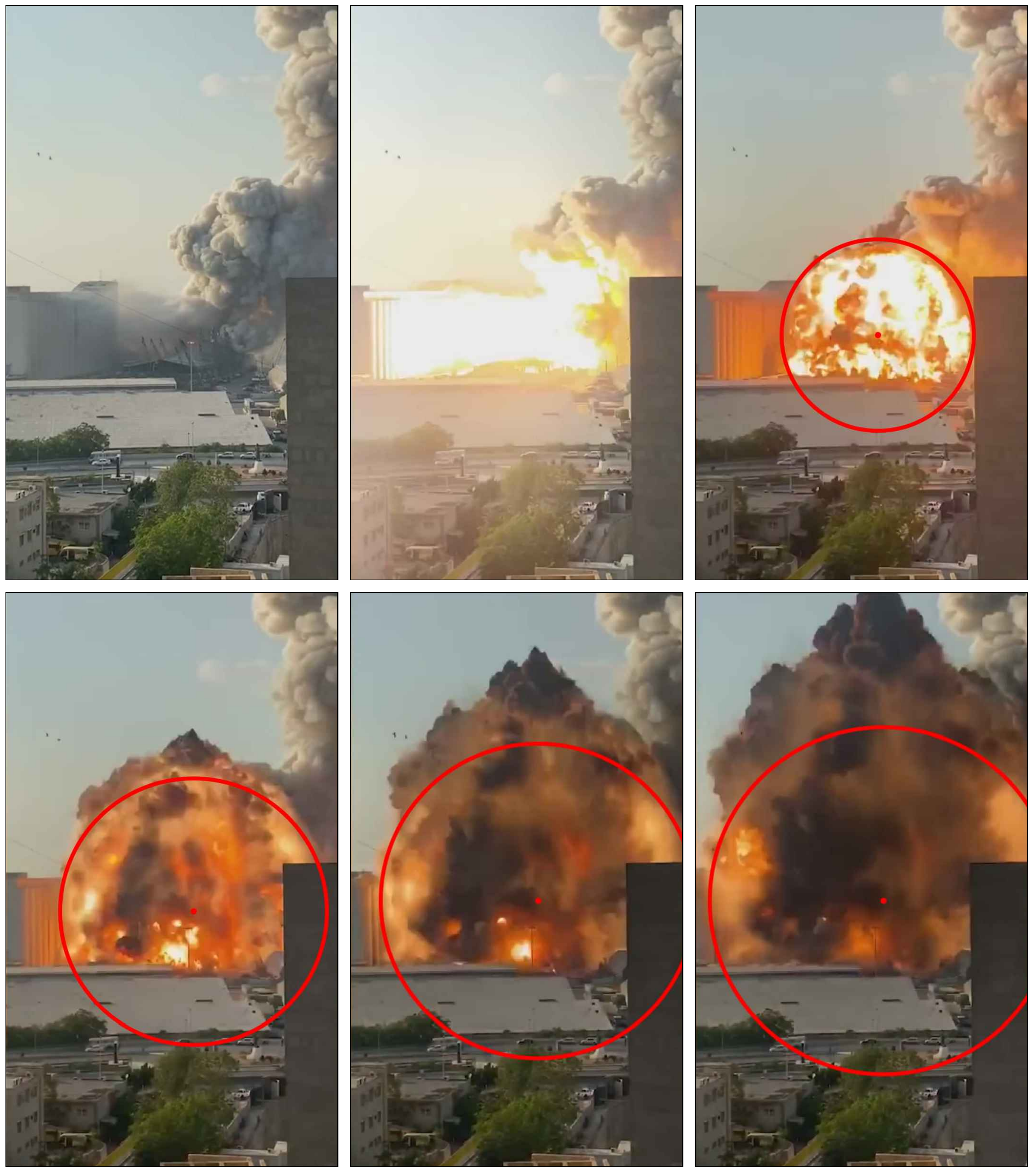}
\caption{Video 3, taken from a distance of 666 meters from the explosion. 6 frames separated by 33.33 ms. The fireball fits are shown in red circles.}
\label{fig:9}       
\end{figure}

\begin{figure*}

\centering
  \includegraphics[width=0.7\textwidth]{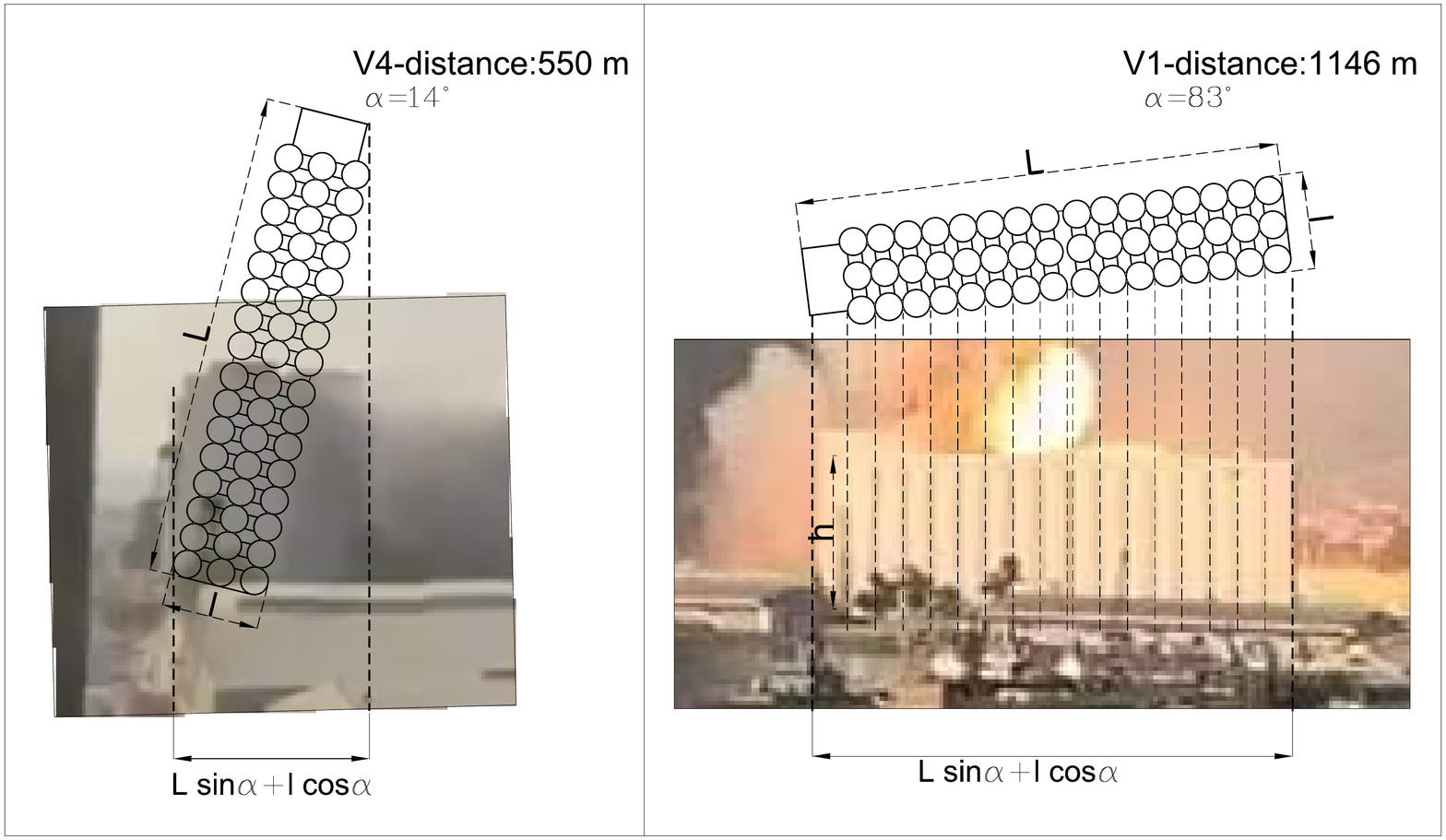}
\caption{For each video the incident angle of sight towards the silos long facade is shown along with the actual picture of the silos and the de-projection lines. Drawings provided by the \cite{aca}. For calibration of video 5 refer to appendix.}
\label{fig:10}       
\end{figure*}
%%%%%%%%%%%%%%%%%%%%
%%%%%%%%%%%%%%%%%%%%%

%
\section{Conclusions and Future Work}
\label{sec:6}
The fireball evolution created by the detonation in Beirut port on the 4th of August 2020 is traced using publicly available videos. The footage are used to extract frames separated by $16.66-33.33$ milliseconds. Pixel calibration is done using the existing silos building and defining accurate line of sight incident angles. Here we draw our main conclusions :
\begin{enumerate}
   
    \item \CA{The evolution of the fireball is used to track the shock front in the first 170 milliseconds until 130 meters from the explosion center, at a scaled distance of $z=1.84 \frac{\mathrm{m}}{\mathrm{kg}^{\frac{1}{3}}}$ in excellent agreement with scaled distances for the shock wave/fireball separation derived from experimental data. The Evolution of the fireball follows a power law with exponent $\approx0.4$ consistent with a Sedov-Taylor model with spherical dimensionality and instantaneous energy release. } 
    \item \CA{The fireball radius evolution follows a theoretical drag model and reaches an asymptotic limit at about $140-145$ meters, a distance beyond which its expansion is brought to a halt.  The scaled distance of this stopping distance using our derived TNT equivalent charge weight $z=2.02 \frac{\mathrm{m}}{\mathrm{kg}^{\frac{1}{3}}}$ is in excellent agreement with other stopping scaled distances observed from published experimental data.}
    \item \CA{We have found a total TNT equivalence of the explosion of $200\pm 80$ tons at a distance range of 60-140 meters in accordance with \textcite{dewey2021tnt}.}
    
    \item \CA{Using a TNT to ammonium nitrate conversion factor of 38\% \parencite{ANequiv2merrifield1991comparison}, we find that the actual mass of ammonium nitrate that participated in the detonation is $526\pm 210 $ tons, much less than the quantity claimed available by the authorities (2750 tons). This can be due either to a late deflagration to detonation transition thus reducing the input in the blast energy, either the missing quantity was not physically available at the time of the explosion.}

    \item  A rapidly expanding vapour cloud has been clearly seen; it is caused by the negative phase after the passage of the pressure wave. Future work shall include observational investigations of its kinematics and possible links to theoretical models.
\end{enumerate}
%%%%%%%%%%%%%%%%%
%%%%%%%%%%%%%%%%%%%
%\section*{Acknowledgments}
\begin{acknowledgements}
The authors would like to thank the referees for their valuable comments which improved the content of the current study. Additionally, the authors express their thoughts to the victims of this tragic event and wishes for the recovery of the wounded. We express our thoughts for the city of Beirut, hoping a prompt and efficient reconstruction.
We primarily thank the eyewitnesses who posted their videos on social media, for without them this article would not have been possible (ex: Elias Abdelnour). We are also thankful to, Alexandra Elkhatib, Samuel Rigby, Jorge Diaz, Gerard-Philippe Zehil, Philip James, Toni el Massih, Nabil Abou Reyali and Igor Chilingarian for useful discussions, and Ranjesh Valavil and Abbas Chamseddine for their help.
\end{acknowledgements}
%\begin{acknowledgements}
%\end{acknowledgements}

 \section*{Conflict of interest}
The authors declare that they have no conflict of interest.
The authors wish to state that referencing a social media profile
does not indicate endorsement of the political or other views of that user.


%%%%%%%%\begin{thebibliography}{}
%\bibliographystyle{ieeetr}
 % \bibliography{ref}
\printbibliography

%\bibitem{ref}
%\bibitem{ref}
% etc
%%%%%%%%%%%%%%\end{thebibliography}
%\clearpage
%\end{multicols}
%
\appendix
\section*{Appendix}
\addcontentsline{toc}{section}{Appendices}
\renewcommand{\thesubsection}{\Alph{subsection}}

%Appendix \arabic{subsection}
\subsection{Calibration of video 5}
%\label{appendix1}

%
Since the silos building is not visible in this video. We locate the position of the camera on the map and determine the distance to a visible face-on a car in the frame. Angular calibration of pixels $\lambda$ (in which a pixel corresponds to an angular measurement) is given by:
\begin{equation} 
\lambda[^{\circ} \mathrm{px}^{-1}]=\frac{2\beta}{26\ \ \mathrm{px}}=\frac{\tan^{-1}{\left(\frac{4.5\ \ \mathrm{m}}{132\ \ \mathrm{m}}\right)}}{26\ \ \mathrm{px}}.
\end{equation}
Here 4.5 m is assumed to be a common length for a car and 132 m and 26 px are the distance to the car and the pixel size of the car respectively.\\
The radius of the fireball in meters $R_{\mathrm{fm}}$ is thus given by
\begin{equation} 
R_{fm}= 630\mathrm{ m } \times \tan\phi=630\mathrm{ m } \times \tan (R_{\mathrm{fpx}} \lambda).
\end{equation}
Here, 630 meters is the distance from the camera to the explosion, and $R_{\mathrm{fpx}}$ is the radius of the fireball in pixels. This is illustrated in Fig. \ref{fig:9}.
We do not perform further corrections due to the deviation of the camera angle with respect to the center of the explosion. These deviations will not cause variations larger than 2 \% on the fireball radius. This is much less than the error margin propagated from all other variables.

\begin{table}[ht]\centering
\caption{Scaling parameters for V5}

\begin{tabular}{lllll}
\hline\noalign{\smallskip}
R &  $\lambda$ & $\tan\phi$ & $R_{fm}$ & T \\
pixels  & $[^{\circ} px^{-1}]$   & & [meters]   &   seconds\\
78 & 0.0765 &0.101 & 65.84 & 0.033 \\
100 & 0.0765 & 0.131 &84.61 & 0.066  \\
120 & 0.0765 & 0.158 &101.81 & 0.099  \\
135 & 0.0765 &0.178 & 114.80 & 0.133  \\
148 & 0.0765 &0.192 &126.13 & 0.166  \\
157 & 0.0765 & 0.208 &134.03 & 0.199  \\
168 & 0.0765 & 0.223 &143.734 & 0.233  \\
\noalign{\smallskip}\hline
\end{tabular}
\end{table}
\begin{figure}[ht]
 \centering
  \includegraphics[width=0.48\textwidth]{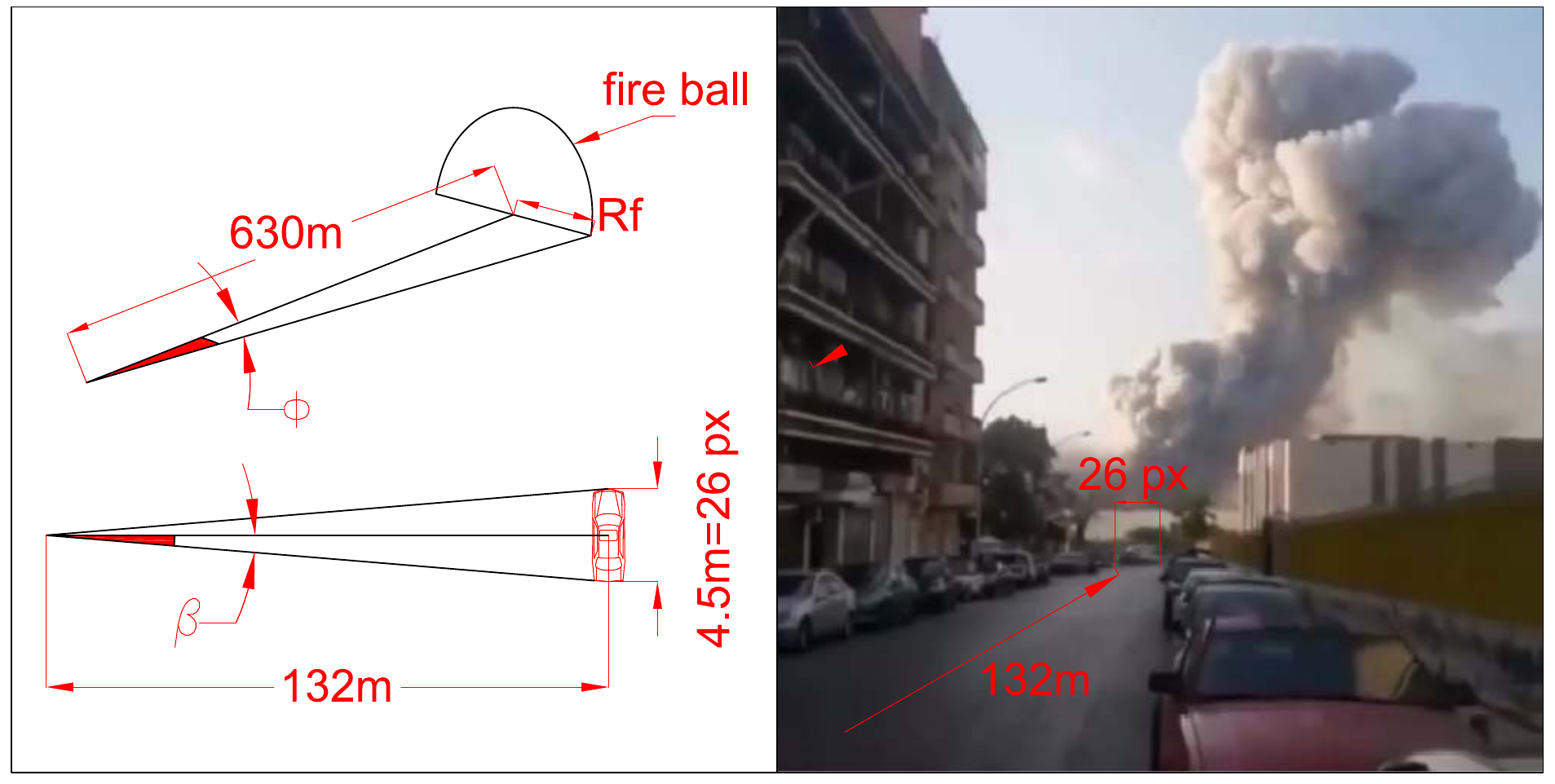}
\caption{Scale calibration for video 5: on the right panel the length of a passing car is measured in pixels. By determining the distance from the general map and assuming a total length of the car of 4.5 meters we calculate the angular field of view per pixel $\lambda$. Using the value of $\lambda$ and knowing the distance to the explosion site, we can measure the angle $\phi$ and convert the pixel measurements of the fireball to meters. }
\label{fig:11}       
\end{figure}
%%%
\subsection{Frames from videos 1-4-6}
\label{appendix2}
Frames extracted from different videos located on the map in fig \ref{fig:2}. For frame rates refer to table \ref{tab:1}. Each frame shows the fireball along with the circle fit to determine the physical length. The detonation is assumed to have happened anytime between the first and the second frame. For pixel calibration method see section \ref{sec:2}.

\begin{figure}[ht]
 \centering
  \includegraphics[width=0.48\textwidth]{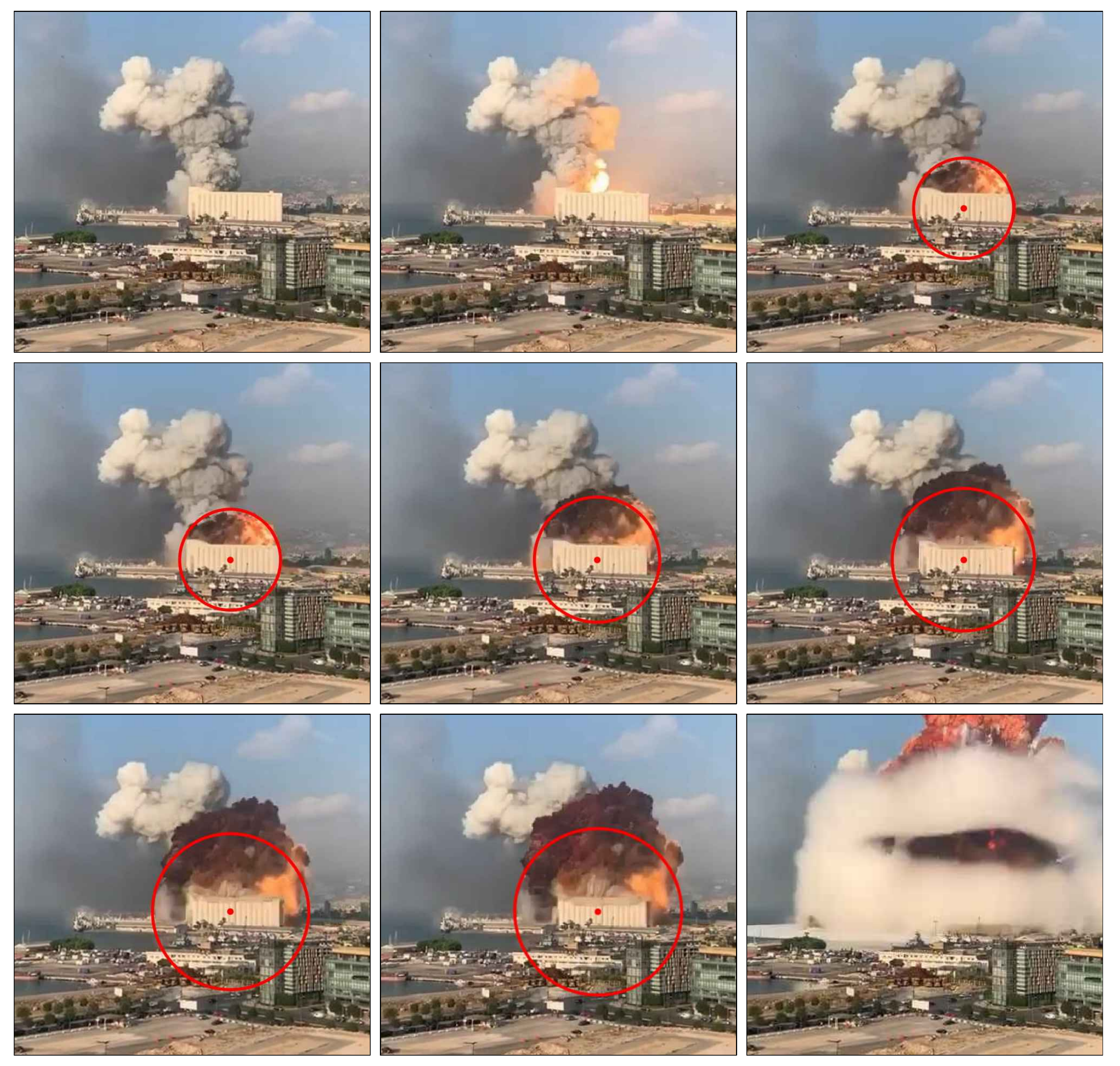}
\caption{Video 1-distance 1146 meters. The dark spot seen within the vapour cloud in the last frame is a trace of the diffraction/interaction of the pressure wave with the existing silos building.}
\label{fig:12}       % Give a unique label
\end{figure}

\begin{figure}[ht]
 \centering
  \includegraphics[width=0.48\textwidth]{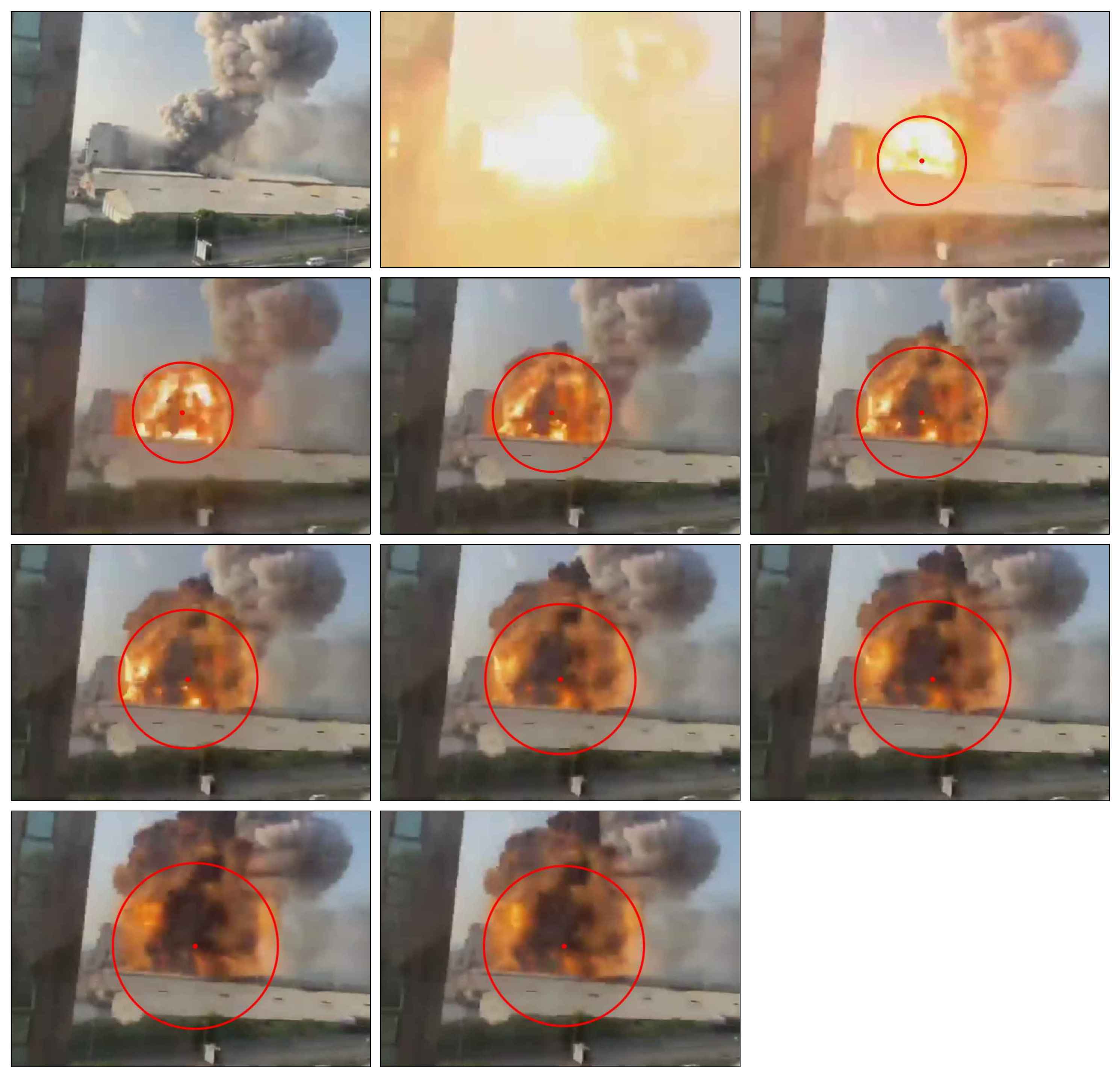}
\caption{Video4-distance 550 m}
\label{fig:13}       
\end{figure}

\begin{figure}[ht]
\centering
  \includegraphics[width=0.48\textwidth]{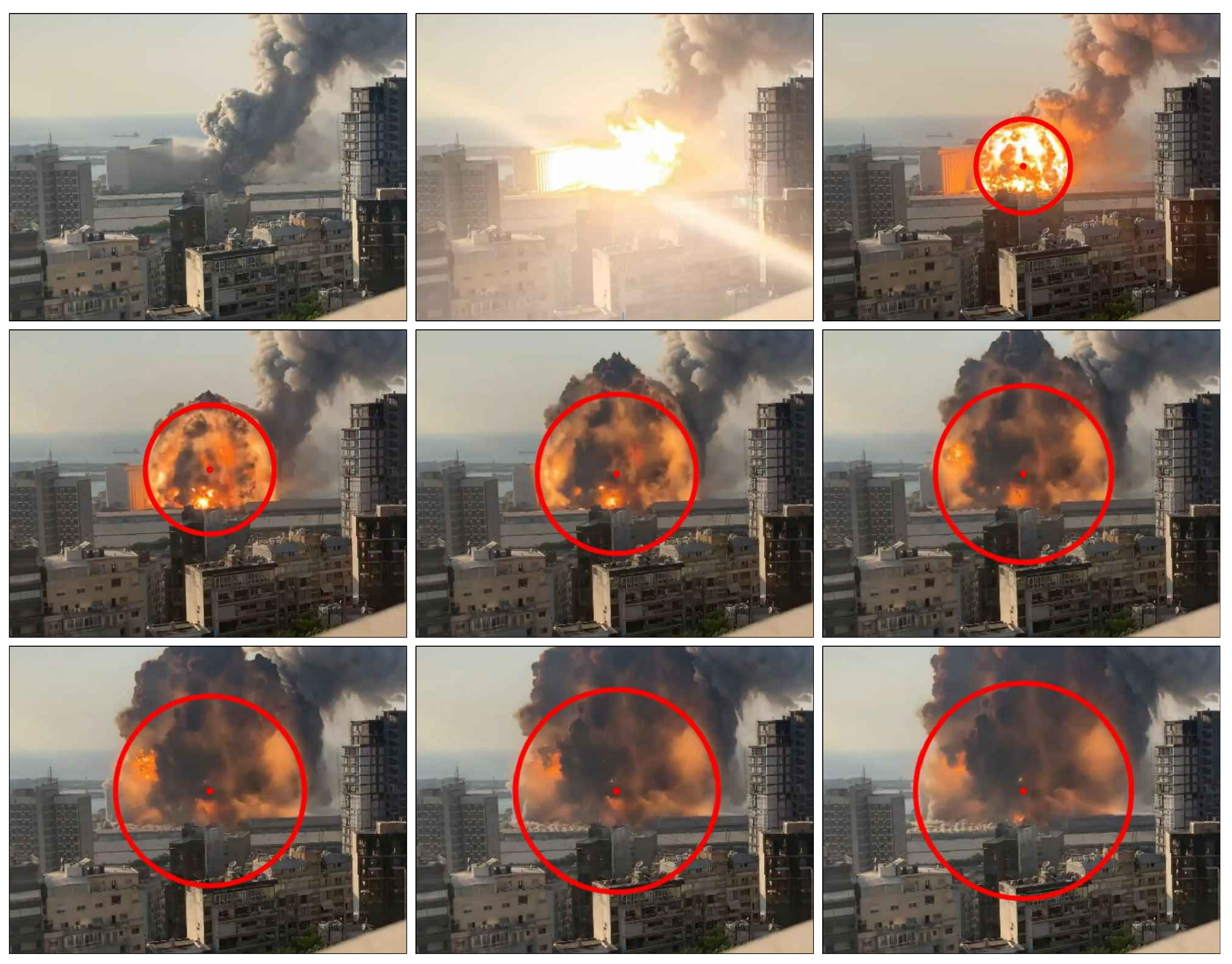}
\caption{Video6-distance 1126 m}
\label{fig:14}       
\end{figure}

%%%%

%%%%%%%%%%%%%%%%%%%%%%%%%
%%%%%%%%%%%%%%%%%%%%%%%%%%
\end{document}